\documentclass[twocolumn,aps,pre,amssymb,amsfonts]{revtex4-1}
\usepackage{amsmath} \usepackage{bm} \usepackage{graphicx}

\begin{document}

\title{Transport in the cat's eye flow on intermediate time scales}

\author{Patrick P\"oschke} 
\affiliation{Institute of Physics, Humboldt University of Berlin, 
Newtonstr. 15, D-12489 Berlin, Germany} 
\author{Igor M. Sokolov} 
\affiliation{Institute of Physics, Humboldt University of Berlin,
Newtonstr. 15, D-12489 Berlin, Germany} 
\author{Michael A. Zaks}
\affiliation{Institute of Physics, Humboldt University of Berlin, 
Newtonstr. 15, D-12489 Berlin, Germany} 
\author{Alexander A. Nepomnyashchy}
\affiliation{Department of Mathematics, Technion, Haifa, 32000 Israel}

\date{\today}

\begin{abstract} 
\noindent We consider the advection-diffusion transport of tracers 
in a one-parameter family of plane  periodic flows where the
patterns of streamlines feature regions of confined circulation 
in the shape of ``cat's eyes'', 
separated by meandering jets with ballistic motion inside them. 
By varying the parameter, we proceed from the regular two-dimensional lattice 
of eddies without jets to the sinusoidally modulated shear flow without eddies.
When a weak thermal noise is added, i.e. at large P\'eclet numbers, 
several intermediate time scales arise, with qualitatively and quantitatively 
different transport properties: 
depending on the parameter of the flow, the initial position of a tracer, 
and the aging time, motion of the tracers
ranges from subdiffusive to super-ballistic.  
Extensive numerical simulations of the aged mean squared displacement for 
different initial conditions are compared to a theory 
based on a L\'evy walk that 
describes the intermediate-time ballistic regime. 
Interplay of the walk process with internal circulation dynamics in the 
trapped state results at intermediate time scale 
in non-monotonic characteristics of aging.
\end{abstract}

\maketitle

\section{Introduction}
In hydrodynamics, the global transport properties of complicated flow patterns 
are derived from (often nontrivial) spatial averages over the local geometry 
of the velocity field. Regions of circulation (eddies, vortices) and the 
far-reaching jets belong to the basic building blocks 
of many two-dimensional flow patterns. 
In laminar jets, the tracer particles are advected over large distances.
In contrast, a tracer captured inside an eddy stays  localized for a long 
time (in the absence of molecular diffusion, forever). 
In 1880, Lord Kelvin (Sir William Thomson) described the inviscid plane 
vortex street flanked by regions of translational motion.  
He portrayed a pattern in which two stripes
with opposite directions of translational velocities were ``separated $[\dots]$
by a cat's eye border pattern of elliptic whirls'' \cite{Kelvin}. In the context
of two-dimensional transport, it is convenient to replace a single vortex street
by a spatially periodic stationary pattern of vortices and jets. 
Below we use for this purpose the so-called ``cat's eye flow'', 
introduced in \cite{Childress1989} for studies of hydromagnetic effects.
This model allows us to vary, by means of the single parameter,
the relative areas occupied by the eddies and by the jets. 
In the limit of shrinking jets, the pattern turns into the pure
cellular flow, i.e. a periodic arrangement
of eddies divided by the separatrices connecting stagnation points.
The separatrices form the cell borders: 
as long as molecular diffusion is neglected, they stay 
impenetrable for the tracers. With diffusion taken into account,
all flow regions become accessible for tracers, and their transport
possesses a hierarchy of timescales: the short and intermediate times 
at which the starting position of a tracer (near the eddy center, close to
a separatrix, etc.) matters, and the final asymptotic state 
in which information about the starting configuration has been effectively
erased by diffusion. For the case of very small molecular diffusion, 
a formal mathematical distinction between intermediate timescales 
was suggested in~\cite{Fannjiang_2002} where transport 
in absence of mean drift  was viewed as
a superposition of an appropriate random continuous martingale process 
and the nearly periodic fluctuation. Two characteristic times were 
introduced: the ``martingale time'', defined as the ratio of the variance 
of fluctuation to the effective diffusivity, and the ``dissipation time'' 
at which the increments of the martingale become approximately
stationary. Over long times, the martingale 
component dominates the behavior: ``the longer the timescale, 
the less anomalous the scaling is''~\cite{Fannjiang_2002}.
Physically, different characteristic times are related to typical
timescales of deterministic circulation as well as to average
durations of diffusive passages across various building
blocks of the flow pattern.

Cellular flows often serve as prime examples of systems
showing subdiffusion for intermediate times, see \cite{Bouchaud, Isichenko} and
references therein. From the point of view of transport, these systems 
have much in common with combs where 
one-dimensional motion along the backbone or spine is interrupted by motion 
along the teeth in the transverse direction. Movement along the 
backbone is modeled by continuous time random walks (CTRW) with power law
waiting time densities. If the stages of propagation along the teeth 
as well as of circulation inside the eddies are regarded as time intervals 
spent in a trapped state, 
then diffusion in cellular flows
corresponds to combs with finite tooth length and to CTRW with power 
law waiting time distributions possessing exponential cutoffs. 
These upper cutoffs originate in the typical maximal time needed to diffuse 
across an eddy. 
Like in other CTRW models with power law waiting times, 
the properties of the diffusion depend on
the aging time and the initial conditions.

In cellular flows with 
weak molecular diffusion,
see \cite{Pomeau, Childress, Soward, Shraiman, RosenbluthEtAl}, 
the intermediate and the final asymptotics are of the main interest. 
The final asymptotics, as predicted by homogenization theory \cite{Majda}, 
is diffusive, and here the recent effort has been
put into the quantitative description. In flows without jets, like in
\cite{Pomeau} or in the eddy lattice flow~\cite{PoeschkeEtAl2016},
the intermediate asymptotics is subdiffusive~\cite{IyerNovikov, HairerEtAl}. 
In flows with jets (``channels'' in terminology
of~\cite{Fannjiang}), this intermediate asymptotics
corresponds to L\'evy walks interrupted by rests. Before addressing
the complex geometry of experimentally available 
flows~\cite{SolomonEtAl1993, SolomonEtAl1994, WeeksEtAl1996},
it seems reasonable to perform a thorough study of a simpler variant:
a two-dimensional periodic flow pattern of the 
``cat's eye flow'' \cite{Kelvin, Childress1989, Fannjiang}.
Below, similar to our previous work on the eddy lattice 
flow \cite{PoeschkeEtAl2016}, we demonstrate that 
for tracers starting on the edge of the jet,
the transport by this flow pattern for intermediate times
can indeed  be modeled as a L\'evy walk (LW), 
with eddies in the role of traps,
and jets viewed as a transport mode in the LW  scheme 
(see \cite{ZaburdaevEtAl} and references therein). 
This conclusion is corroborated by comparison of theoretical estimates 
with results of our extensive numerical simulations. However, due to the
presence of internal circulation dynamics in the trapped state, 
the model based on the flow
possesses a richer behavior than a simple L\'{e}vy walk model would suggest, 
which is manifested in its very different aging properties.
In contrast to the well understood aging in LW~\cite{Barkai,Magdziarz},
the cat's eye flow displays strong dependence on the initial conditions 
and a set of unusual aging behaviors. 
Like in the eddy lattice, transport by the flow
is more complicated than its random walk representation! We are aware neither
of existing extensive numerical simulations of this system, 
nor of a comparison of numerics with theoretical predictions.

Below, in Sec.~\ref{sect_Model} we discuss and illustrate the basic features
of the cat's eye flow pattern. The subsequent
Sec.~\ref{sect_MSD} focuses on the different time scales 
for the mean squared displacement of tracer particles in the system. 
In Sec.~\ref{sect_Numerics} we present and discuss the results of 
numerical simulations. 
Finally, in Sec.~\ref{sect_Conclusion} we sum up our findings. Some details
of the theoretical description are contained in the appendix.

\section{The flow} 
\label{sect_Model}

Dynamics of a tracer in the plane cat's eye flow obeys
the stochastic differential equation
\begin{equation} 
\dot{\mathbf{r}} = \mathrm{rot}\,
(0,0,\psi_{\mathrm{Cat}}(\mathbf{r}))+\sqrt{2D}\bm{\xi} 
\end{equation} 
with the two-dimensional~\cite{footnote1} stream function 
\begin{equation}
\label{original}
\psi_{\mathrm{Cat}}(x,y) = u\,a\left( \sin\left( \frac{x}{a}\right)\sin\left(
\frac{y}{a}\right) +A\cos\left( \frac{x}{a}\right)\cos\left(
\frac{y}{a}\right)\right), 
\end{equation} 
where $u$ is the characteristic
velocity, $D$ is the molecular diffusivity and $\bm{\xi}=(\xi_x, \xi_y)$ 
is a vector of Gaussian noises with zero mean, unit variance and 
$\langle \xi_x(t) \xi_x(t')\rangle = \langle \xi_y(t) 
\xi_y(t')\rangle =\delta(t'-t)$. 
The deterministic part of the flow pattern
is periodic with respect to both coordinates and consists of elementary cells 
of the length and width $\pi a$. On taking $a$ as the spatial unit,
${a^2}/{D}$ as the unit of time, and introducing the
P\'eclet number $\mathrm{Pe}={u\,a}/{D}$, the equations turn into
\begin{eqnarray} 
\label{Langevin} 
\dot{\mathbf{r}} &=& \mathrm{Pe}\;
\mathrm{rot}\, (0,0,\Psi_{\mathrm{Cat}}(\mathbf{r}))+\sqrt{2}\bm{\xi} \\
\label{psi} \Psi_{\mathrm{Cat}}(x,y) &=& \sin x \sin y +A\cos x \cos  y.
\end{eqnarray} 
Thus the system is governed by two dimensionless parameters
$A$ and Pe. We concentrate on the case $\mathrm{Pe}\gg 1$.
The deterministic velocity components of the flow
(\ref{psi}) are 
\begin{eqnarray}
\label{DiffEq} 
\mathrm{Pe}^{-1}\dot{x} &=& 
\frac{\partial \Psi_{\mathrm{Cat}}}{\partial y} 
= \sin x \cos y - A \cos x \sin y \\ \nonumber
\mathrm{Pe}^{-1}\dot{y} &=& -\frac{\partial \Psi_{\mathrm{Cat}}}
{\partial x} = -\cos x \sin y +
A \sin x \cos y. 
\end{eqnarray} 
This flow pattern can be imposed in a layer of incompressible fluid 
with kinematic viscosity $\nu$ that obeys the Navier-Stokes equation,
by applying a spatially periodic force, 
e.g. $\mathbf{F} = 4\nu \sin x \cos y\,(\mathbf{e}_x + A\mathbf{e}_y)$. 

Formally, the parameter $A$ can assume arbitrary real values of
either sign. However, it is sufficient to restrict analysis to the interval 
$0 \leq A \leq 1$. A transformation $x \to \pi - x$ (or $y \to \pi - y$) is
equivalent to the change of the sign of $A$, whereas a shift $x \to x + \pi/2$,
$y \to y + \pi/2$ with simultaneous rescaling of time units by the factor $A$
is equivalent to the transformation $A \to 1/A$. 
For numerical investigations, we take the following values of $A$:
$10^{-3},\,10^{-2},\,10^{-1},\,0.25,\,0.5,\,0.75,\,0.9,$ and $1$.

\begin{figure*}[t] \centering
\includegraphics[width=80mm]{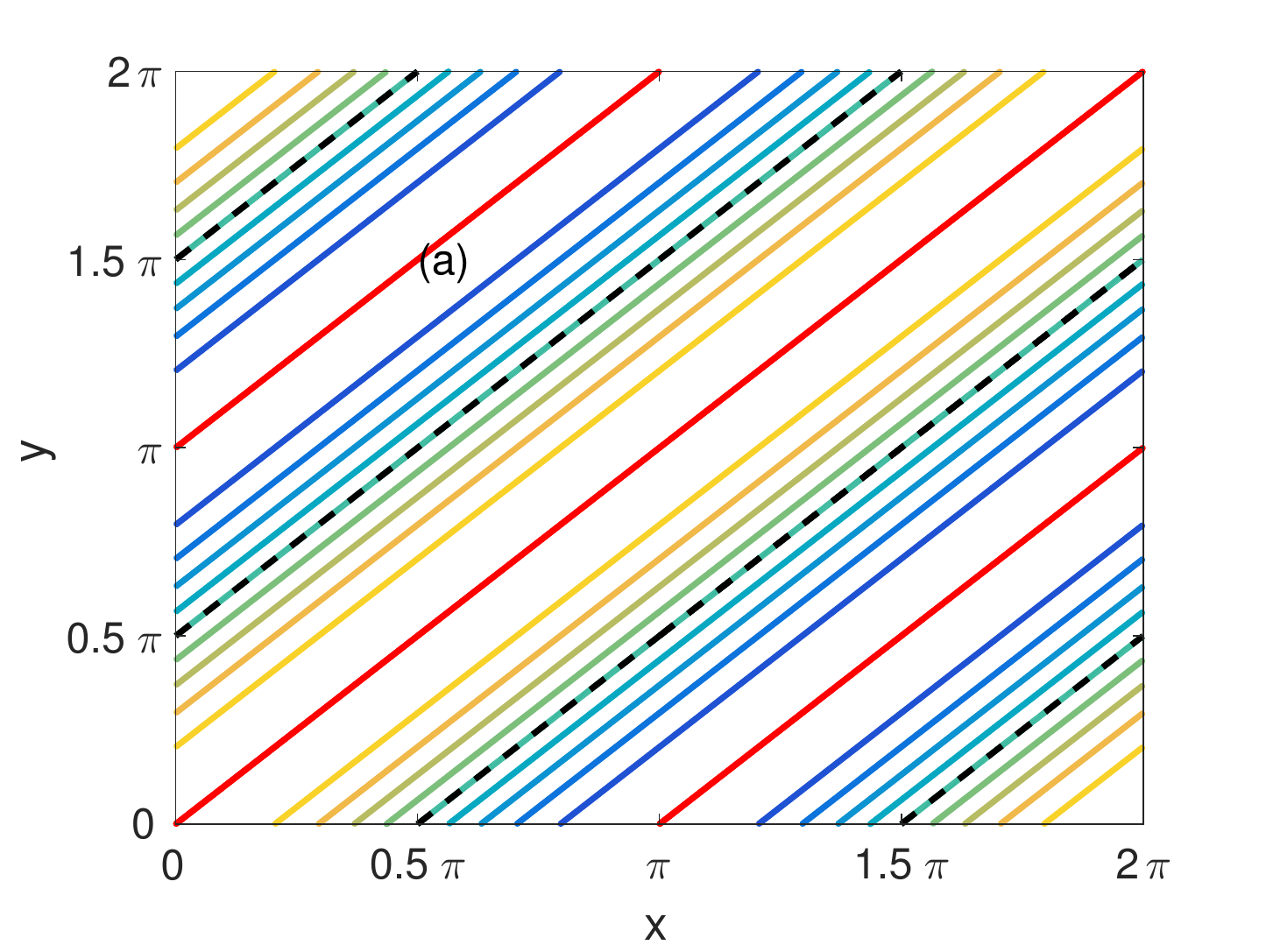}
\includegraphics[width=80mm]{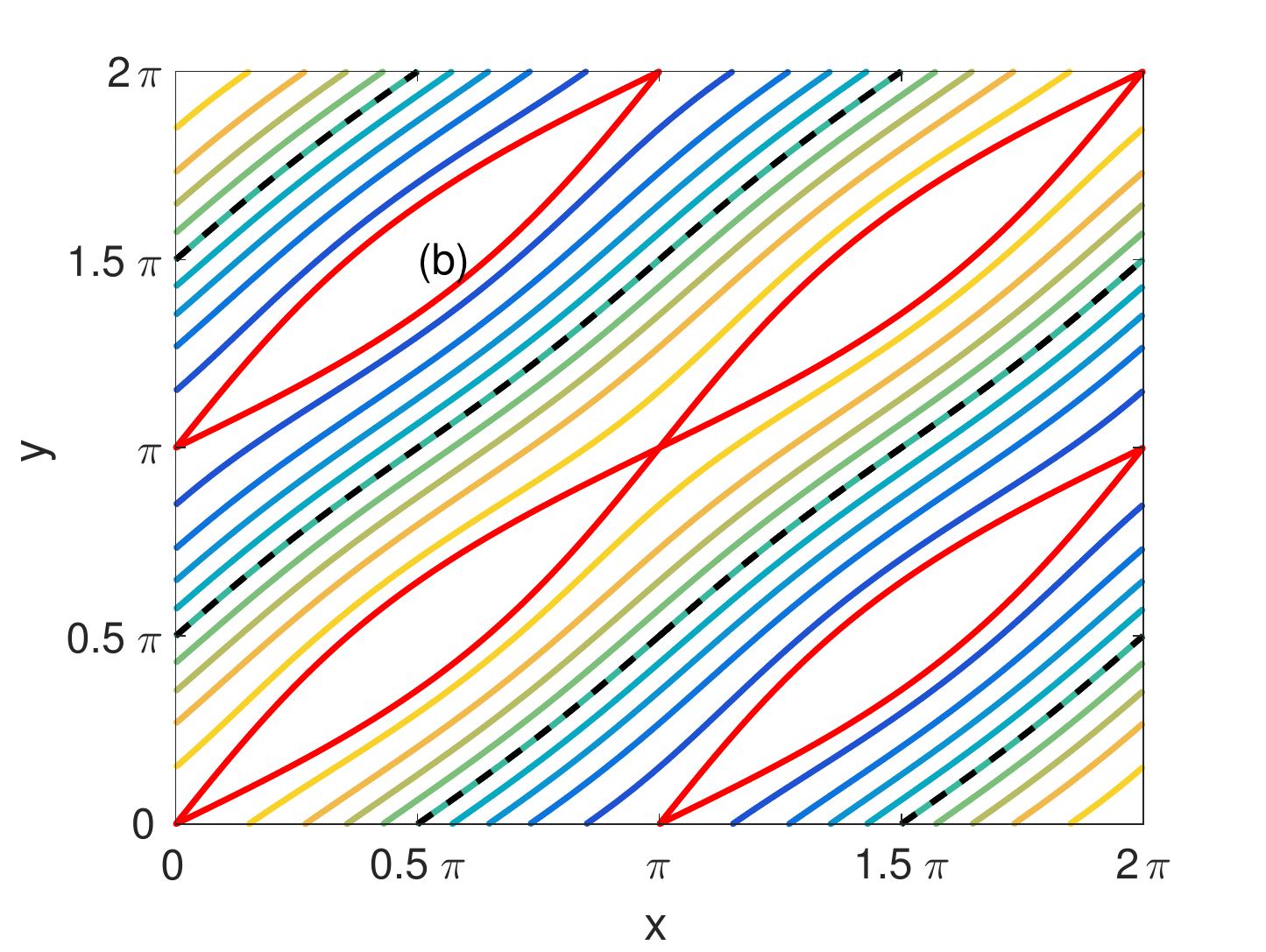}
\includegraphics[width=80mm]{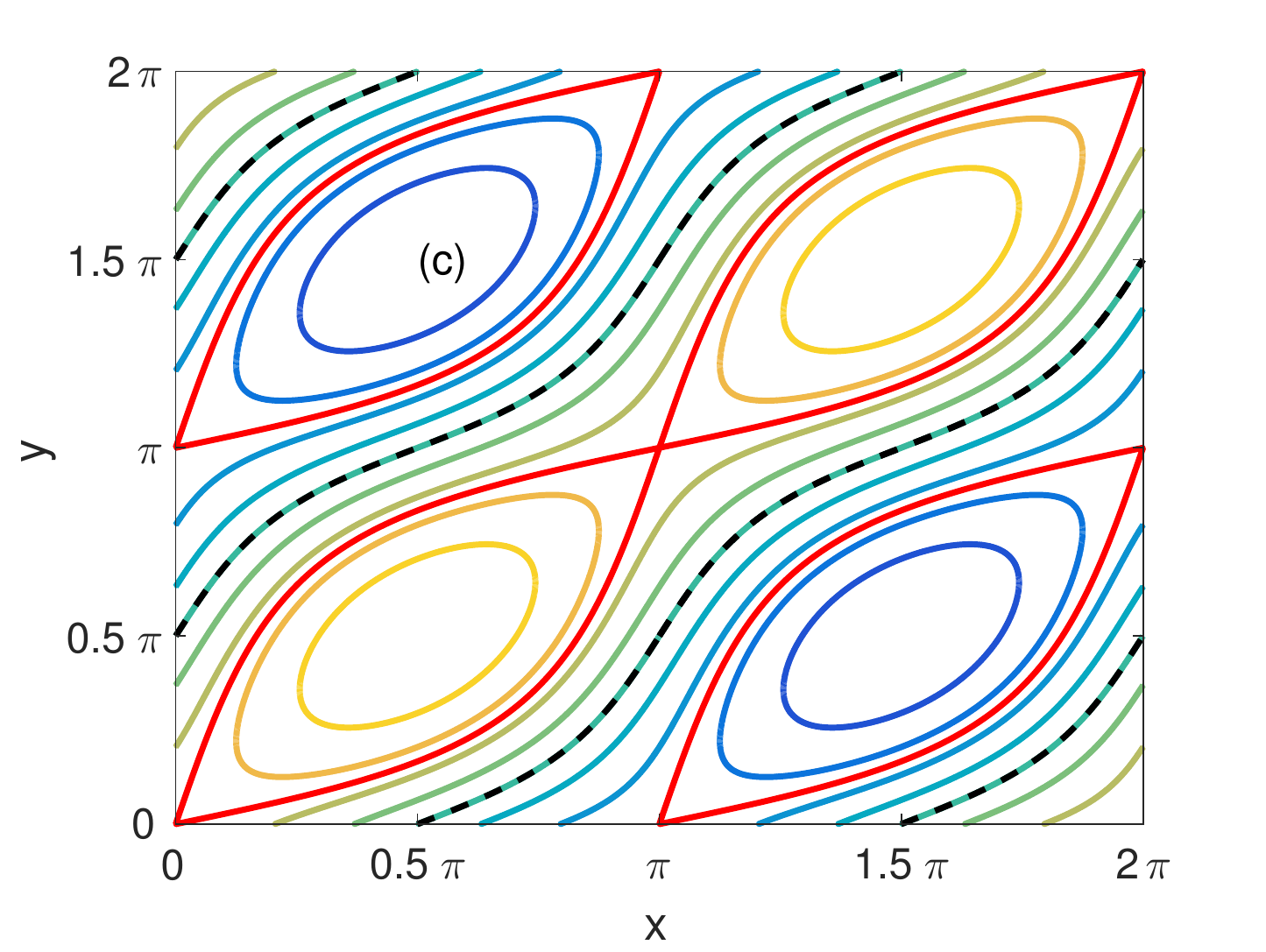}
\includegraphics[width=80mm]{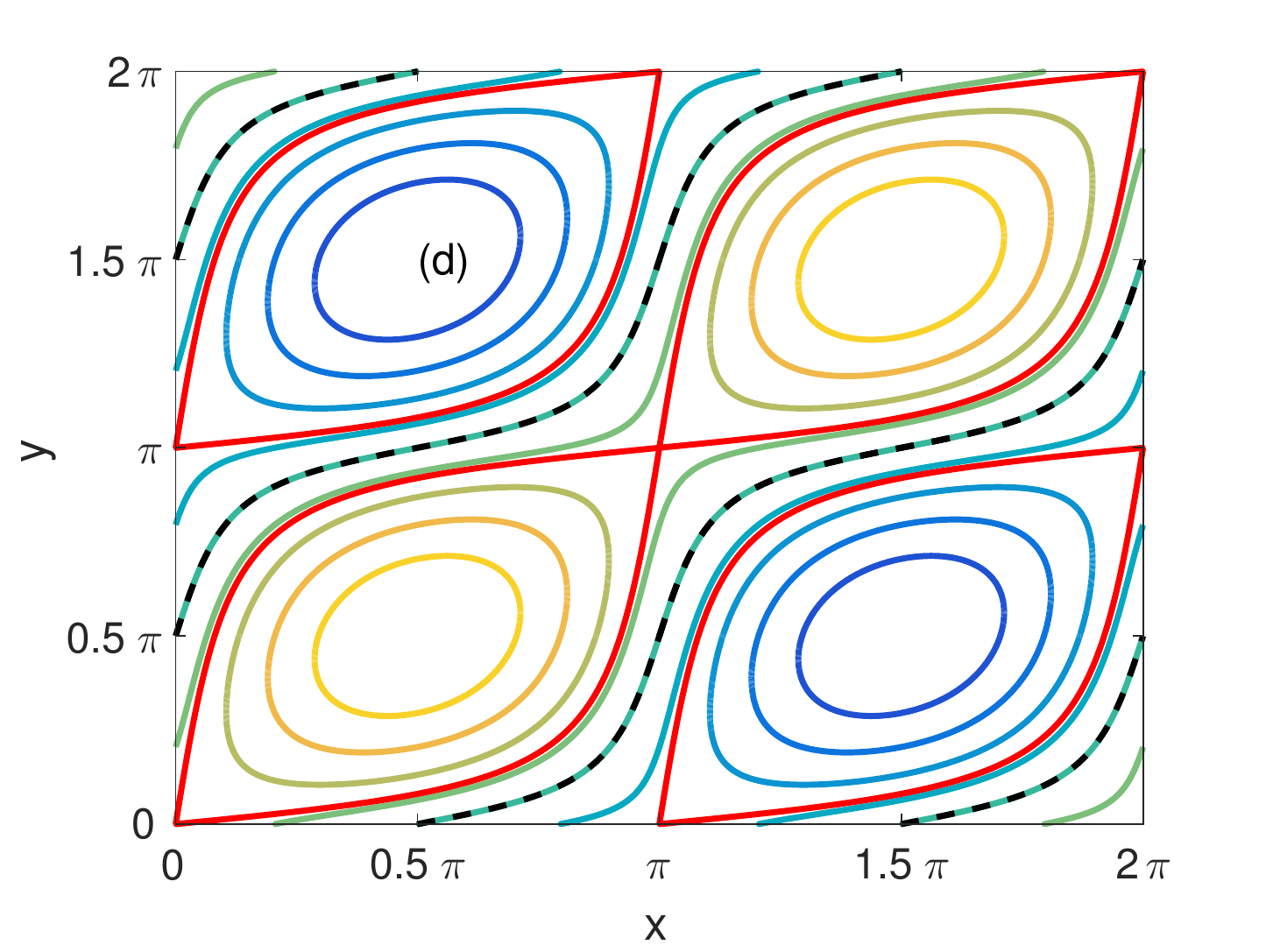}
\includegraphics[width=80mm]{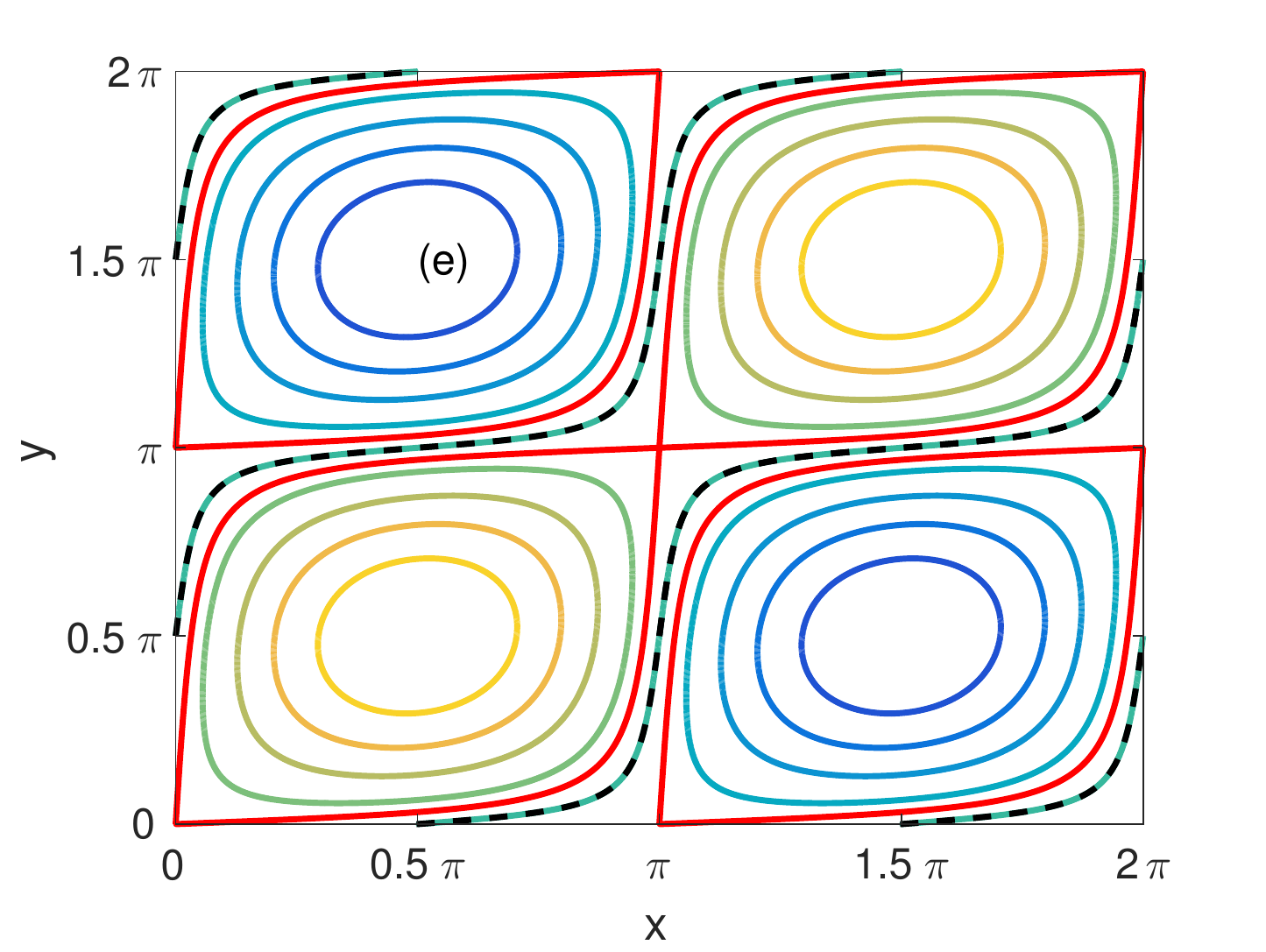}
\includegraphics[width=80mm]{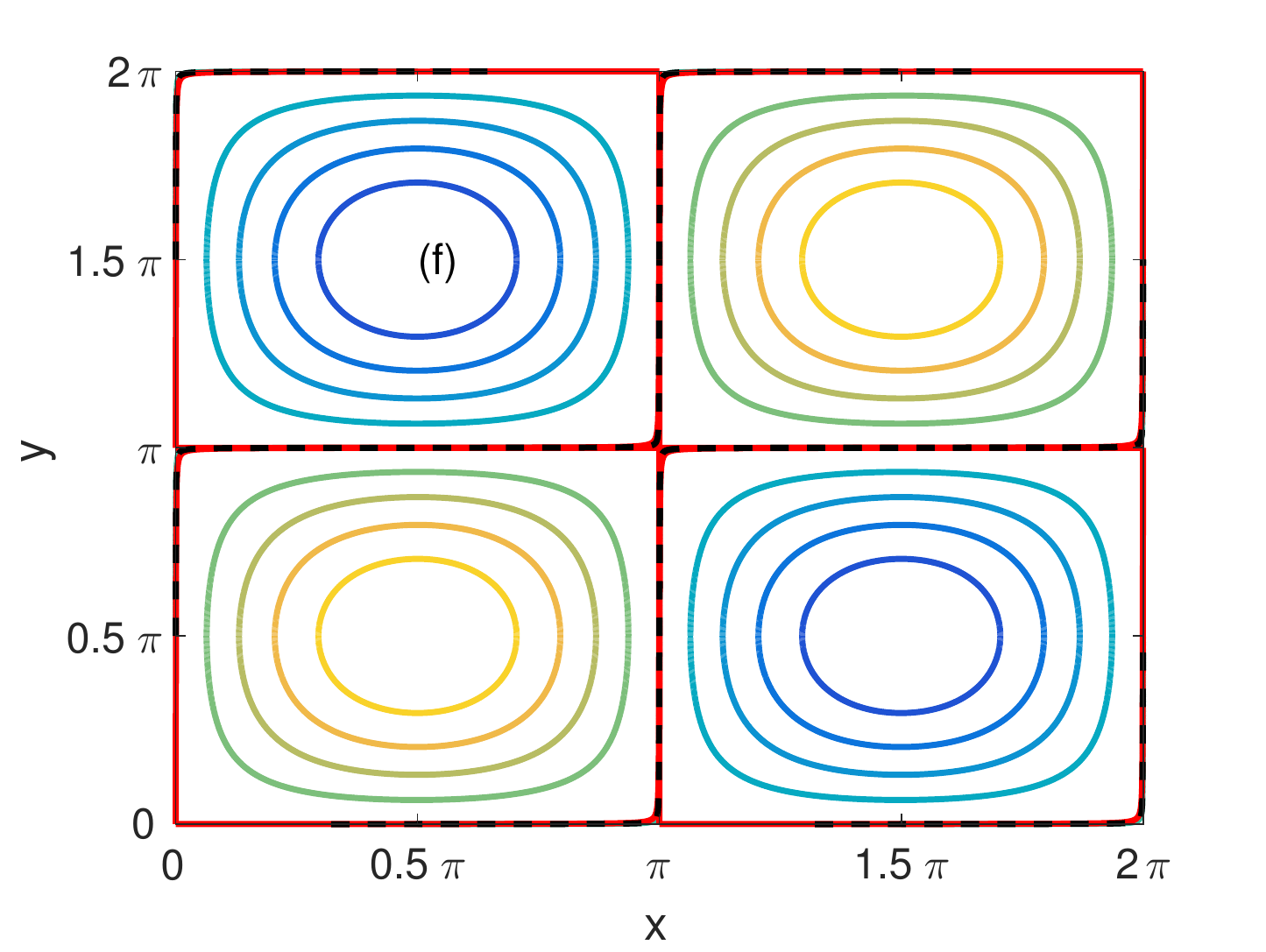}
\caption{\label{contour}Contour plot of streamfunction (\ref{psi}) 
for (a) $A=1$, (b) $A=0.9$, (c) $A=0.5$, (d) $A=0.25$, (e) $A=10^{-1}$, 
and (f) $A=10^{-3}$. 
Yellow closed streamlines, e.g. lower left vortex, denote 
counterclockwise motion. Blue closed streamlines denote clockwise motion. 
For $A\to 1$ separatrices (red) between jets and eddies merge pairwise, 
the eddies cease to exist, and a shear flow with the sinusoidal velocity 
profile emerges.
For $A\to 0$ pairs of separatrices (red) merge with the midline of the jet
(black dashed) and become the edges of quadratic cells.} \end{figure*}

Regardless of the value of $A$, the flow possesses stagnation
points at ($x = \pi m$, $y = \pi n$),  and at ($x
= \pi/2 + \pi m$, $y = \pi/2 + \pi n$),
$m, n = 0, \pm 1, \pm2, \dots$. At $|A| < 1$, the former points are
hyperbolic fixed points (saddles) and the latter ones are elliptic fixed
points (centers). At $|A| > 1$, the reverse configuration of equilibria
takes place. Exchange of stability between stagnation points occurs in the
course of degenerate global bifurcation at $|A| = 1$. At this parameter value,
the straight lines $y = x+\pi n$, $n = \pm 1, \pm 2 \dots$ turn into
invariant  continua of stagnation points, see straight red lines 
in Fig.~\ref{contour}~(a).
In that case the system is transformed into a shear flow:
the plane is  partitioned into alternating regions of ballistic
motion in opposite directions. At $A=0$, in contrast, the jets are absent
and the entire plane is covered by cells with closed streamlines:
the eddy lattice flow~\cite{PoeschkeEtAl2016}. 
Isolines of the stream function (\ref{psi}) for several typical values of $A$
are presented in Fig.~\ref{contour}. The dashed curves, obtained by shifting 
by multiples of $\pi$ in both directions the curve
\begin{equation} 
\label{jetcenter} 
y(x) = \arctan\left(-A\cot(x)\right), 
\end{equation} 
for $x\in[0, \pi]$, 
are the midlines of the jet regions in which
ballistic motion takes place: along them, $\Psi_{\mathrm{Cat}}(x,y)$ vanishes.
These jet regions are separated from the closed elliptic orbits 
by the isolines $|\Psi_{\mathrm{Cat}}|=A$: 
the separatrices are obtained by translating  
\begin{equation}
\label{separatrix} y_{\pm}(x) =
\arccos\left[
\frac{A^2\cos x\pm\sqrt{1-A^2}\sin^2 x}{A^2\cos^2 x +\sin^2 x }\right]
\end{equation} 
for $x\in[0, \pi]$ along both coordinates with
$\pi$-periodicity, cf. the red curves that delineate cat's eyes 
in Fig.~\ref{contour}. Here the plus (respectively minus) sign
corresponds to the lower (respectively upper) boundary of the ``cat's eye''.
At non-zero small values of $A$ the narrow curvy jets with
alternating directions of unbounded motion are formed between the cells
(Fig.~\ref{contour}f). As $A$ grows, these jets become thicker
(Fig.~\ref{contour}e, Fig.~\ref{contour}d).

The flow is anisotropic: it possesses an axis of faster
transport (shortened to ``the axis'' throughout this paper). 
Its direction corresponds to rotation of the
coordinate system $(x,y)$ around the origin by $\pi/4$.
In the rotated reference frame, the equations of motion are simplified:
in terms of $x_{\pm} = x \pm y$, their deterministic part turns into
\begin{eqnarray} 
\label{DiffEqR} 
\dot{x}_+ &=& \mathrm{Pe}\,(1+A)\sin x_-,\nonumber\\ 
\dot{x}_- &=& \mathrm{Pe}\,(1-A)\sin x_+. 
\end{eqnarray} 
At $A = 1$ the value of $x_-$ becomes an integral of motion, 
and the plane gets foliated into the continuum of invariant straight
lines.  The velocity of motion along each of these lines is sinusoidally
modulated across the continuum. In presence of diffusion, the (longitudinal)
motion along $x_+$ has both deterministic and diffusive components, whereas the
(transverse) motion along $x_-$ is purely diffusive.  
In contrast, at $A = 0$, the pattern (\ref{DiffEqR}) turns into the 
conventional cellular flow. In terms of
$x_+$ and $x_-$, the separatrix (\ref{separatrix}) becomes
\begin{equation} 
x_- = \pm\arccos\left[\frac{(1-A)\cos x_+ + 2A}{1+A}\right],
\end{equation}
hence the maximal width of the ``eye'', in terms of original coordinates
$x$ and $y$, is
$$ \sqrt{2}\arccos \frac{3A-1}{1+A}.$$
The local velocity $v_{\rm s}$ along the separatrix 
is given by
\begin{eqnarray}
\label{vel_sep}
v_{\rm s}^2(x_+) &=& 2\mathrm{Pe}^2(1-A) \\
\nonumber &&
\times\left[(A-1)\cos^2 x_+ -2A\cos x_+ +A+1\right].
\end{eqnarray}

We take for the width of the jet channel the distance between the separatrices of adjacent saddle points,
measured along the local normal direction to the midline $\Psi_{\mathrm{Cat}}=0$ of the jet, yielding a lengthy expression.
As a function of the coordinate $x_+$, 
along the axis,
the jet width oscillates between the sharp maximum
\begin{equation}
\label{max_width}
w_{\rm max} =\frac{1}{\sqrt{2}}\arccos \frac{1-3A}{1+A}
\end{equation}
and the minimal value
\begin{equation}
\label{min_width}
w_{\rm min} =\frac{1+A}{\sqrt{1+A^2}}\arcsin\frac{2A}{1+A}
\end{equation}
measured in units of the original coordinates. Both reproduce the exact width $\pi/\sqrt{2}$ of the jet for $A=1$.

At small values of $A$ the width displays a broad plateau around its
minimal value. 
There, the minimum 
\begin{equation}
\label{width_typical}
w_{\rm min}\approx 2A+\frac{A^3}{3}+\ldots
\end{equation}
can be used as a ``typical'' jet width. The linear approximation suffices for our purposes. 
At $A=1$ the deviation is 22\%. For $A\leq 0.9$ it is 4\% or less, fitting better for smaller values of $A$.

\section{Mean squared displacement} 
\label{sect_MSD}

\subsection{Characteristic times}

We start from the analysis of the characteristic times involved in the
motion of the particles. 
The treatment is analogous to the case of eddy 
lattices~\cite{PoeschkeEtAl2016}, and the notation used is similar. 
In comparison to the eddy lattices
with their two characteristic times:
\begin {itemize} 
\item $t_1$ - the
characteristic time of the deterministic transport over a single eddy, 
\item $t_2$ - the characteristic time to diffuse through a cell of
the size $\pi a$,
\end {itemize} 
here the third time,
\begin {itemize} 
\item $t_3$ - the characteristic maximal time spent in a jet,
\end {itemize} 
comes into play, making the picture more complex. 

The estimates for the two first times are the same as in the eddy lattice flow, 
see \cite{PoeschkeEtAl2016}. Expressed in units of, respectively, 
Eq.(\ref{original}) and Eq.(\ref{Langevin}),  they are 
\begin{equation} 
t_1 \simeq \frac{a}{u} =\frac{1}{\mathrm{Pe}}
\label{t1} 
\end{equation} 
and 
\begin{equation}
t_2 \simeq \frac{a^2}{D} \equiv 1,
\label{t2} 
\end{equation} 
and are interrelated  via the P\'eclet number: 
$t_1=t_2/\mathrm{Pe}$.  The third characteristic time is 
\begin{equation} 
t_3 \simeq \frac{a^2}{D} w^2 = t_2 w^2
\end{equation} 
where $w$ is the characteristic width of the jet measured in units of $a$, 
i.e. the parameter $w$ itself is dimensionless. 
Note that
\begin{equation} 
\label{t3} t_3 = w^2 \approx 4A^2 
\end{equation} 
in our normalised units.

The waiting times in an eddy are given by a power law probability density
function (normal Sparre-Andersen behavior)
\begin{equation} 
\phi(t) \propto t^{-3/2} 
\end{equation}
between $t_1$ and $t_2$ with
cutoffs both at short and at long times. The waiting time in a jet is given by 
a similar power law, but with the upper cutoff time $t_3$. The time $t_1$
corresponds essentially to the time resolution of the random walk scheme: 
behavior at shorter times is dominated by the local dynamics. 

\subsection{Transport regimes}

When starting at the separatrix of the flow, three transport regimes can be
observed. Here we describe them qualitatively; quantitative details are
relegated to the subsequent section 
that presents the numerical simulations of the motion.

At short times $t < t_1$, a particle starting anywhere close to the
separatrix, except for the immediate vicinity of the hyperbolic stagnation
point, moves along the streamline 
with the local velocity close to $v_s$ 
and the average velocity $v$
of the order of Pe. 
This motion does not depend on whether the instantaneous
position of the tracer is inside the jet or inside the eddy. The regime 
of motion is therefore ballistic: 
The mean squared displacement MSD of a particle grows as 
$\langle (\Delta R)^2 \rangle \simeq v^2t^2$. 
The simulations confirm that at large P\'eclet numbers the
typical velocities 
when moving close to the separatrix in the jet 
and in the eddy coincide, and are 
close to Pe: 
$v \approx \mathrm{Pe}$. 

\begin{figure}[h] 
\centering
\includegraphics[width=80mm]{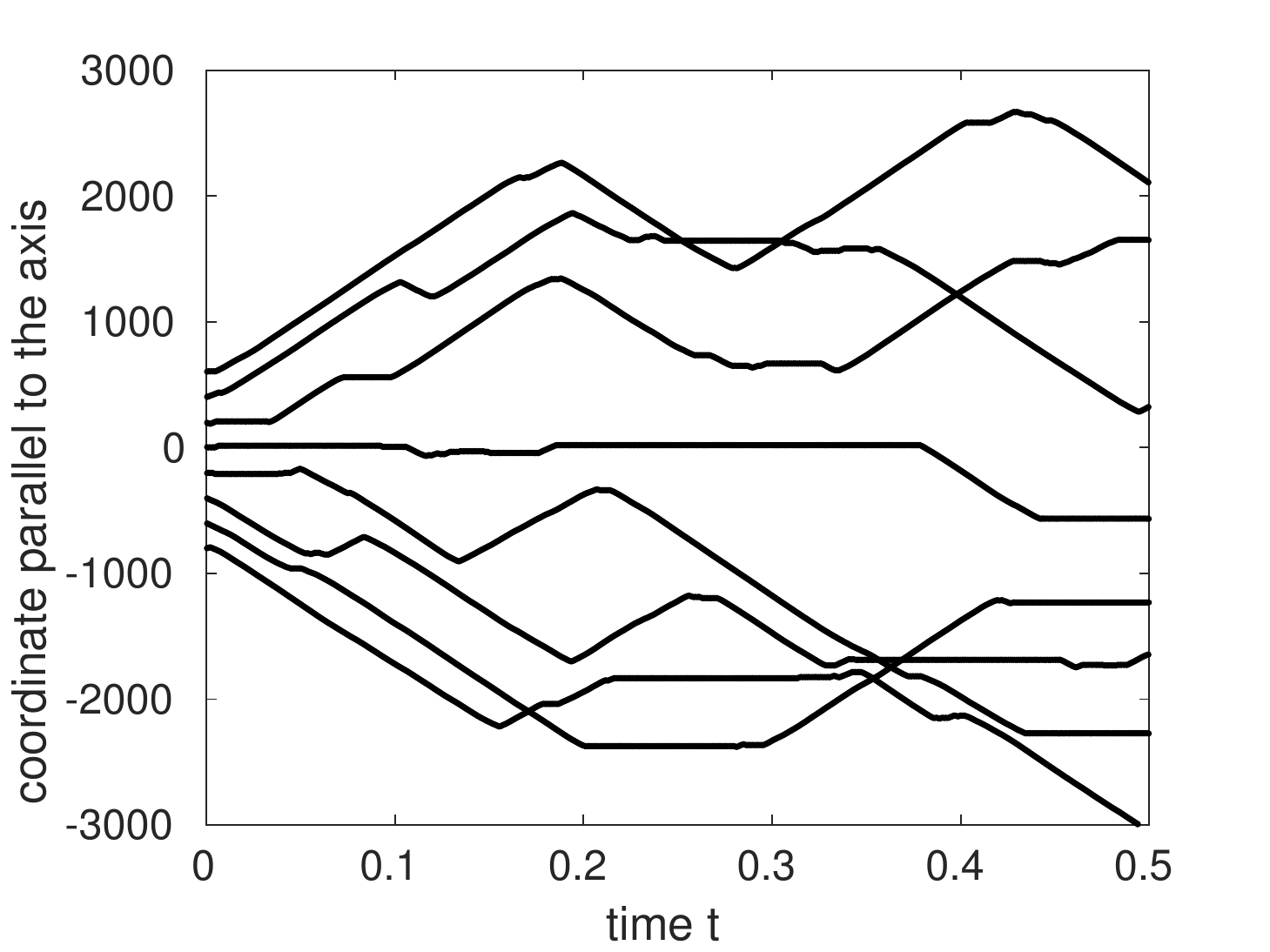}
\caption{\label{traj}
Coordinate parallel to the axis of the system at
$\mathrm{Pe}=10^4$ and $A=0.5$ for a tracer starting at the separatrix. 
To resolve the trajectories optically at small values of $t$, 
a fictitious vertical shift between them has been introduced.} 
\end{figure}

\begin{figure*}[t] \centering
\includegraphics[width=80mm]{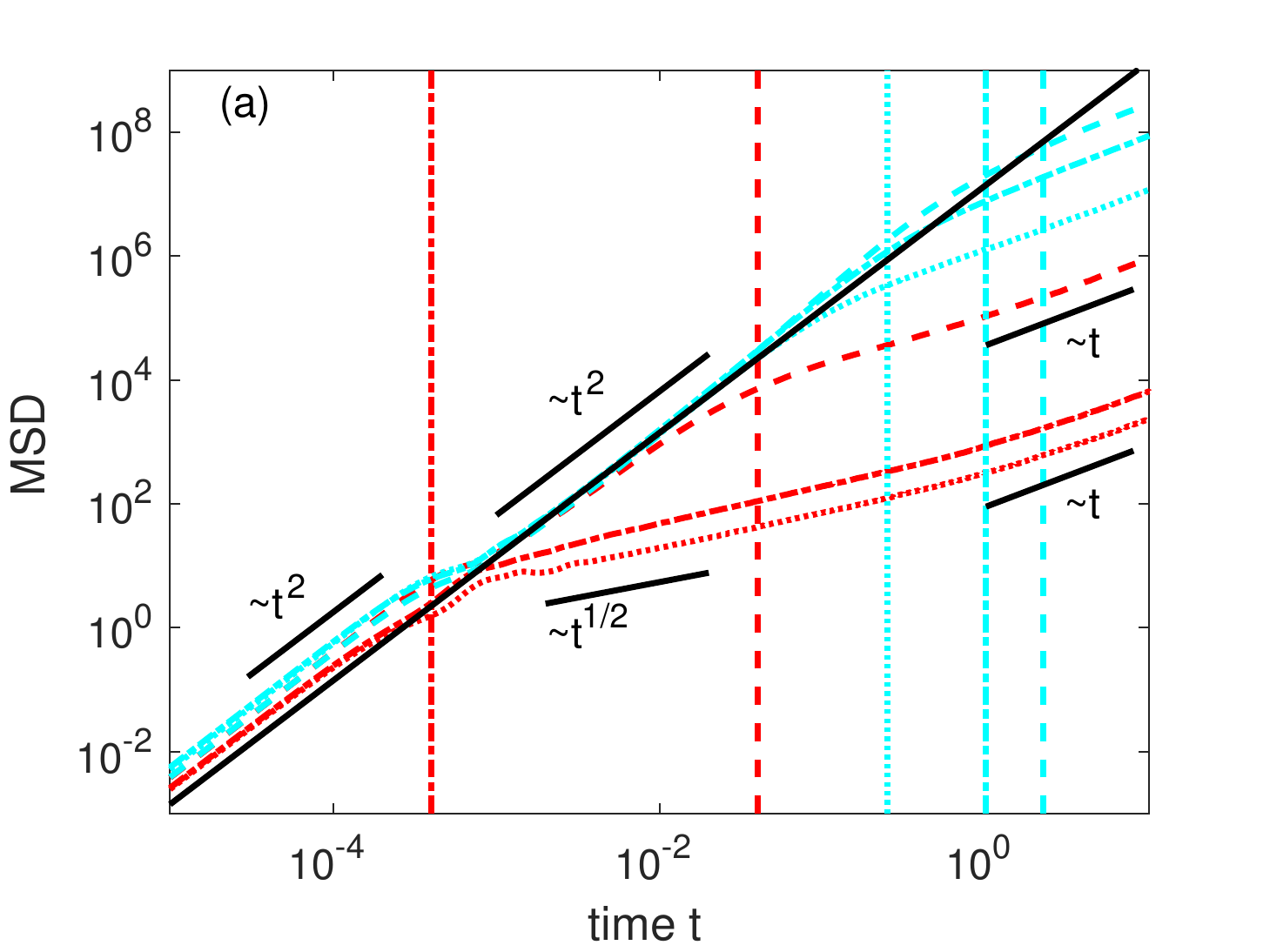}
\includegraphics[width=80mm]{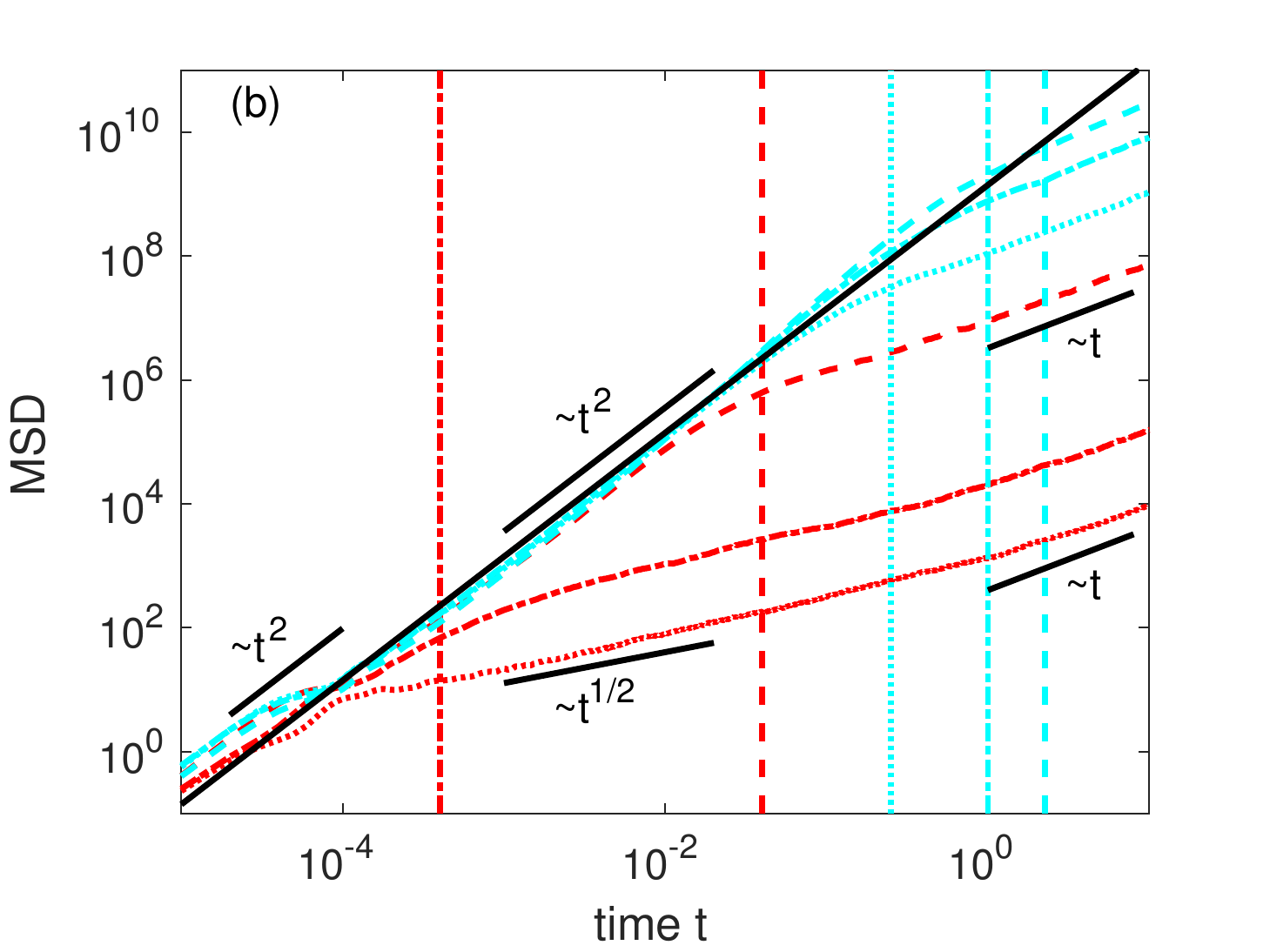}
\caption{\label{MSDsep} Temporal evolution of MSD for ensembles of tracers starting
on the separatrix between the jet and the vortex 
at, respectively,  $A=10^{-3}$ (lowest curve),
$A=10^{-2}$, $A=10^{-1}$, $A=0.25$, $A=0.5$ (center curves),
and $A=0.75$ (uppermost curve), 
compared to the asymptotic theory, Eq.~(\ref{Second_ballistic})
(black continuous). Time $t_3$ Eq.~(\ref{t3})
is indicated by vertical lines.
Panel (a): $10^4$ walks at Pe $=10^4$.
Panel (b): $10^3$ walks at Pe $=10^5$.
Velocity in both plots varies slightly with $A$ around $v\approx 0.75\, \mathrm{Pe}$. 
The intermediate ballistic regime occurs only if $t_3\gg t_1=1/\mathrm{Pe}$.
Note that the curves in (a) are the black dashed lines 
in Figs.~\ref{MSD1} and \ref{MSDlast}.} 
\end{figure*}

\begin{figure}[h] 
\centering
\includegraphics[width=80mm]{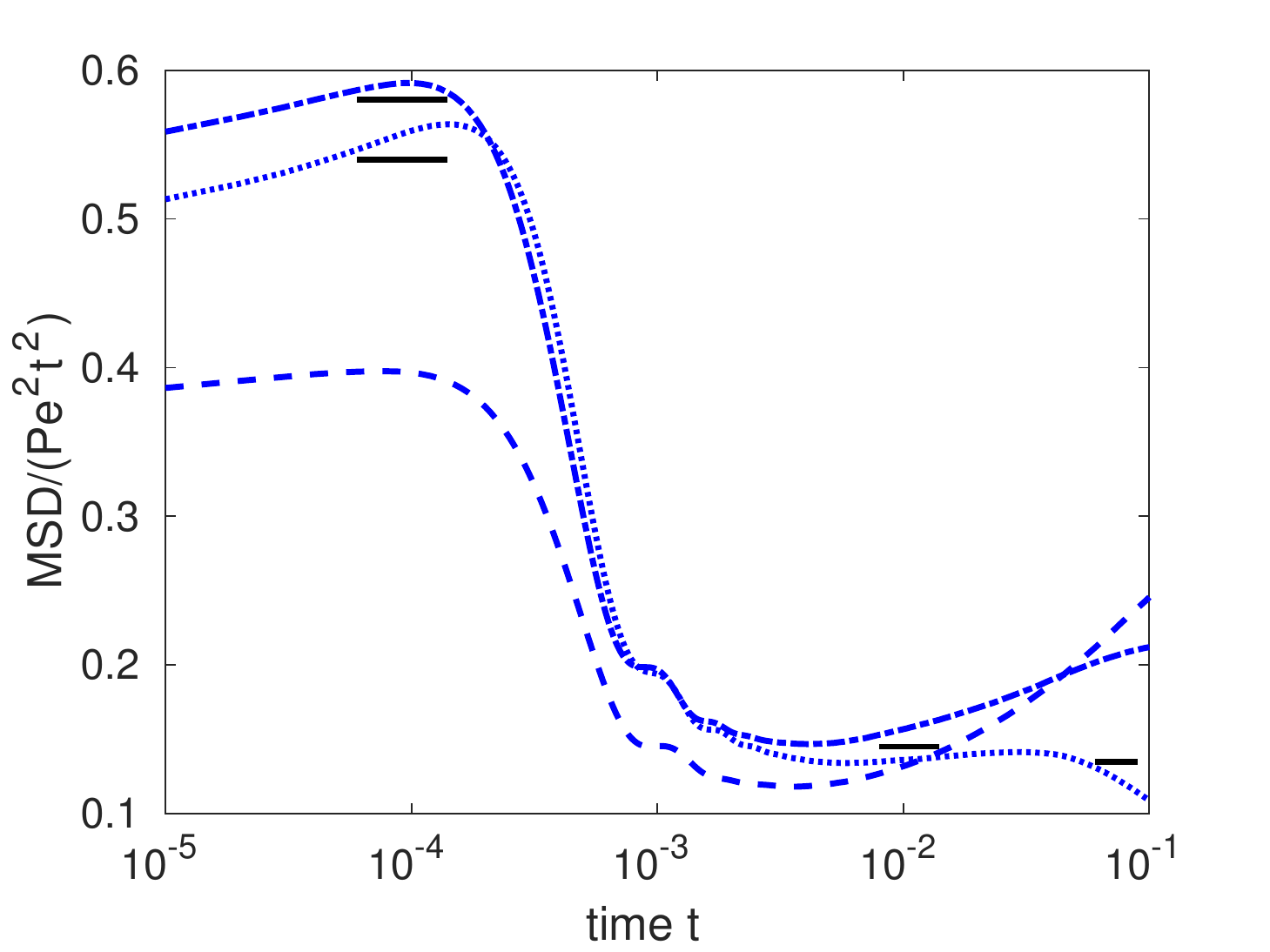}
\caption{\label{MSDsepResc} Crossover in prefactor of MSD between two regimes of ballistic transport.
Rescaled curves from Fig.~\ref{MSDsep} (a) for $A=0.25$ (dotted), 
$A=0.5$ (dash-dotted), and $A=0.75$ (dashed). 
Horizontal lines indicate the values: $0.58$, $0.54$, $0.145$, and $0.135$. 
The MSD during the intermediate ballistic regime is about one quarter of 
its value during the initial ballistic regime, 
as predicted by theory, Eq.~(\ref{Second_ballistic}).} 
\end{figure}

After the time $t_1$, provided both times $t_2$ and $t_3$ are considerably
larger than $t_1$, i.e. $\mathrm{Pe}\,w^2 \gg 1$, the second transport regime 
sets in. Like a preceding regime, this is still a ballistic
transport, but a slower one: the prefactor is reduced to a quarter of
its original size. 
At these times the particle might travel over several lattice cells. 
The translational
motion in the jet corresponds to the transport mode of a L\'evy walk scheme,
whereas the confined motion in an eddy corresponds to the trapping event. 
Both stages 
can be visually identified in the trajectories in Fig.~\ref{traj}. 
The waiting time densities in the trapped and
in the transport modes follow the same power-law asymptotics 
$\phi(t) \propto t^{-3/2}$. 
Such a L\'evy walk scheme corresponds to the ballistic motion 
and the analysis within the L\'evy walk formalism 
(see Appendix~\ref{sect_B}) yields the expression for the growth of MSD: 
\begin{equation}
\label{Second_ballistic}
\mathrm{MSD}(t) = \frac{1}{4}v^2 t^2.
\end{equation}   
Subsequent regimes of ballistic transport for various 
values of $A$ are presented in Fig.~\ref{MSDsep}.
The crossover between them is visualized in Fig.~\ref{MSDsepResc},
where it can be seen that the theoretical estimate for the change
in the prefactor,  given by Eq.~(\ref{Second_ballistic}),
is well matched by simulations.
The MSD is dominated by the
longitudinal motion along the axis of the system. 
The (anomalous) transport in
the system is strongly anisotropic since the motion in direction normal to the
axis still takes place within just one eddy or jet. This motion cannot be 
captured by a coarse-grained random walk model, and will be discussed 
further on the basis of numerical results. 

The behavior between the two larger times $t_3$ and $t_2$ is non-universal. 
In contrast, the terminal diffusion regime that sets in at
long times $t > t_2$ is universal: it does not depend on initial conditions. 
In the remainder of this section we reproduce the estimates for
directional coefficients of diffusion in the terminal regime, 
derived by Fannjiang and Papanicolaou~\cite{Fannjiang}.

At time and length scales corresponding to the terminal regime, 
the flow structure can be considered as a layered one,
a parallel arrangement of jets and rows of eddies 
whose particular form practically ceases to play a role:
the terminal diffusion coefficients are dominated by the times spent in the
corresponding structures. A simple estimate is based on the
assumption of constant thickness for those parallel layers: the exact form 
and the thickness modulation influence only numerical prefactors. 
For small $A$ the eddy lattice (EL) layers show isotropic diffusion 
with the diffusion coefficient 
\begin{equation}
D_{\mathrm{EL}} \simeq \sqrt{u a D} = D\, \mathrm{Pe}^{1/2}. 
\end{equation} 
The diffusion in the channel (jet) is strongly anisotropic. 
In the direction normal to the axis 
\begin{equation} 
D_\bot= D 
\end{equation} 
holds, whereas in the parallel direction we have 
\begin{equation} 
D_\| \simeq u^2 t_3 = u^2 \frac{a^2 w^2}{D} = D\, \mathrm{Pe}^2 w^2.
\end{equation} 

The terminal diffusion coefficient in the direction perpendicular to the axis 
is given by the harmonic mean of the corresponding local coefficients, i.e. 
it corresponds to a sequential switching of diffusivities 
(conductivities in electric terms) 
\begin{equation} 
D^*_\bot \simeq (D_{\mathrm{EL}}^{-1} (1-w) + 
       D_\bot^{-1} w)^{-1} = \frac{D}{w + (1-w) \mathrm{Pe}^{-1/2}}.  
\end{equation}
For large $\mathrm{Pe}$ it is dominated by the first term in the denominator
\begin{equation}
D^*_\bot \simeq D/w \simeq D/ A .
\end{equation} 
This effect takes place for $w > \mathrm{Pe}^{-1/2}$, which in
the units $u=a=1$, used in \cite{Fannjiang}, translates precisely into
$A > \sqrt{D}$. The final diffusion coefficient in the direction parallel to 
the axis is the one for parallel switching of the diffusivities (conductivities)
\begin{equation} D^*_\| \simeq D_{\mathrm{EL}} (1-w) + D_\| w = D
[\mathrm{Pe}^2 w^3 + \mathrm{Pe}^{1/2} (1-w)]. 
\end{equation} 
For large $\mathrm{Pe}$ this is dominated by the first term provided 
$w >\mathrm{Pe}^{-1/2}$ again, resulting in 
\begin{equation} D^*_\|\simeq D \mathrm{Pe}^2 w^3, 
\end{equation} i.e. 
\begin{equation} D^*_\| \simeq D
\mathrm{Pe}^2 A^3.
\end{equation} 
In this way,
the result of \cite{Fannjiang} is reproduced;
since there $\mathrm{Pe} \equiv 1/D$, it follows $D^*_\| \simeq A^3/D$. 
 
The simple reasoning above does not allow us to analyze the
aging phenomena that strongly depend on the nontrivial dynamics inside the
flow structures. This task can be accomplished numerically.

\section{Numerical simulations} 
\label{sect_Numerics}

By numerically integrating the Langevin equation
(\ref{Langevin}), we obtained trajectories $\mathbf{r}(t)=(x,y)$ of tracer
particles. These trajectories were used to compute the 
MSD from the initial position, 
as well as the MSD parallel respectively
perpendicular to the axis. Integration was performed by  
the stochastic Heun method: an efficient algorithm for integration 
of stochastic differential equations with additive noise~\cite{Mannella}. 
The size of the time step was chosen sufficiently small to ensure that
the deviations of the deterministic part of (\ref{Langevin}) from its exact
solution are negligible.  Note that without noise the
stream function $\Psi$ is conserved. Choosing a time step 
$\Delta t=10^{-3}\mathrm{Pe}^{-1}$ turned out to be sufficient 
for all parameter values in all regimes of interest, until the maximum
simulation time $t_{\mathrm{max}}=10$. The simulations were done 
in the range of P\'eclet numbers from $10^{3}$ to $10^5$. 
Below, we focus mainly on the results for $\mathrm{Pe}=10^4$.

We consider the following four initial conditions: 
(a) starting at the center of a cat's eye, i.e. at $(x,y)=(\pi/2, \pi/2)$, 
(b) ``flooded'', i.e. with equal probability 
in the periodic cell of the two-dimensional space, 
(c) starting on the central streamline of the jet, 
i.e. with $x$ drawn with equal probability from $[0,\pi]$ 
and obeying (\ref{jetcenter}), and 
(d) starting from the separatrix between jet and cat's eye, i.e. 
with $x\in[0, \pi]$  equally distributed and obeying Eq.~(\ref{separatrix}). 
Results for different values of $A$ are plotted in
Figs.~\ref{MSDsep} to \ref{MSDlast}. For most situations the component of the 
MSD parallel to the axis dominates. Only for small values of $A$ or 
when starting inside the cat's eye for short time intervals the two 
components are approximately equal.

\subsection{Starting from the separatrix}

When starting from the separatrix the MSD shows an initial and an intermediate
ballistic regime, see Fig.~\ref{MSDsep}. One can clearly see, that the
intermediate ballistic regime occurs only if $t_3\gg t_1=1/\mathrm{Pe}$.

In the other case, when $A$ is very small so that the flow pattern
is very close to the cellular flow, the intermediate
diffusion exponent is 1/2 \cite{IyerNovikov, HairerEtAl}.

\subsection{Other initial conditions}

\begin{figure*}[t] 
\centering
\includegraphics[width=80mm]{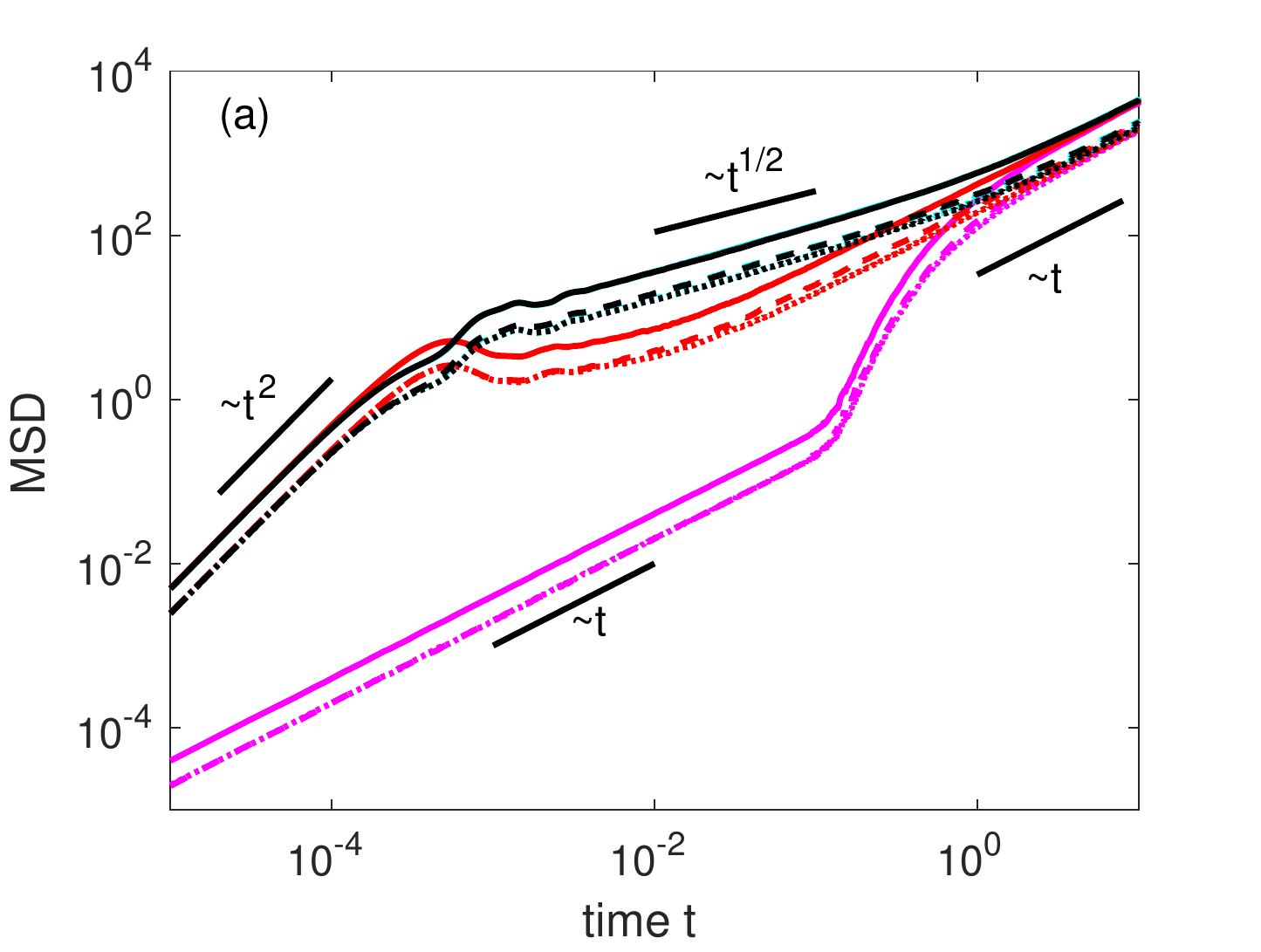}
\includegraphics[width=80mm]{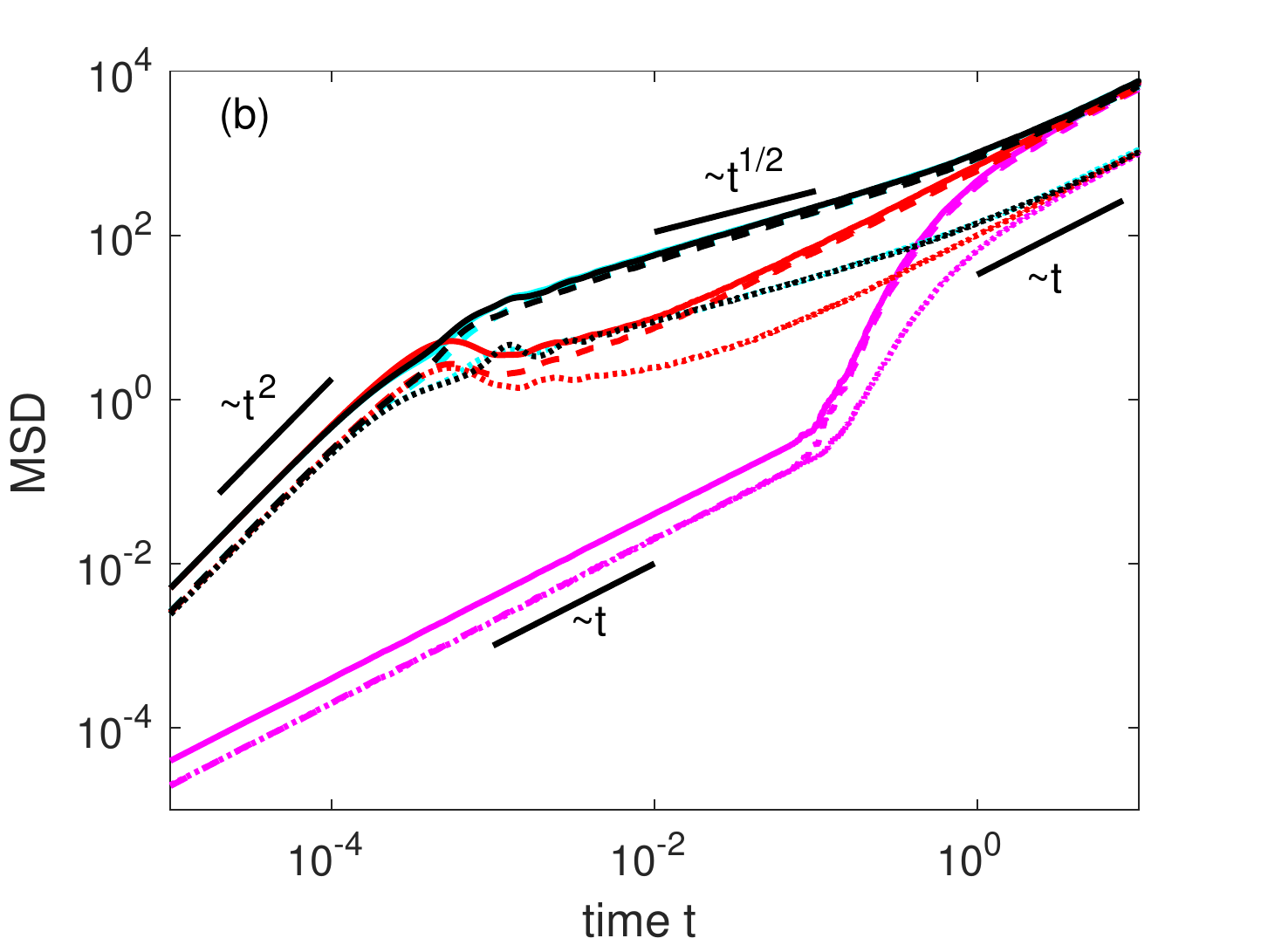}
\includegraphics[width=80mm]{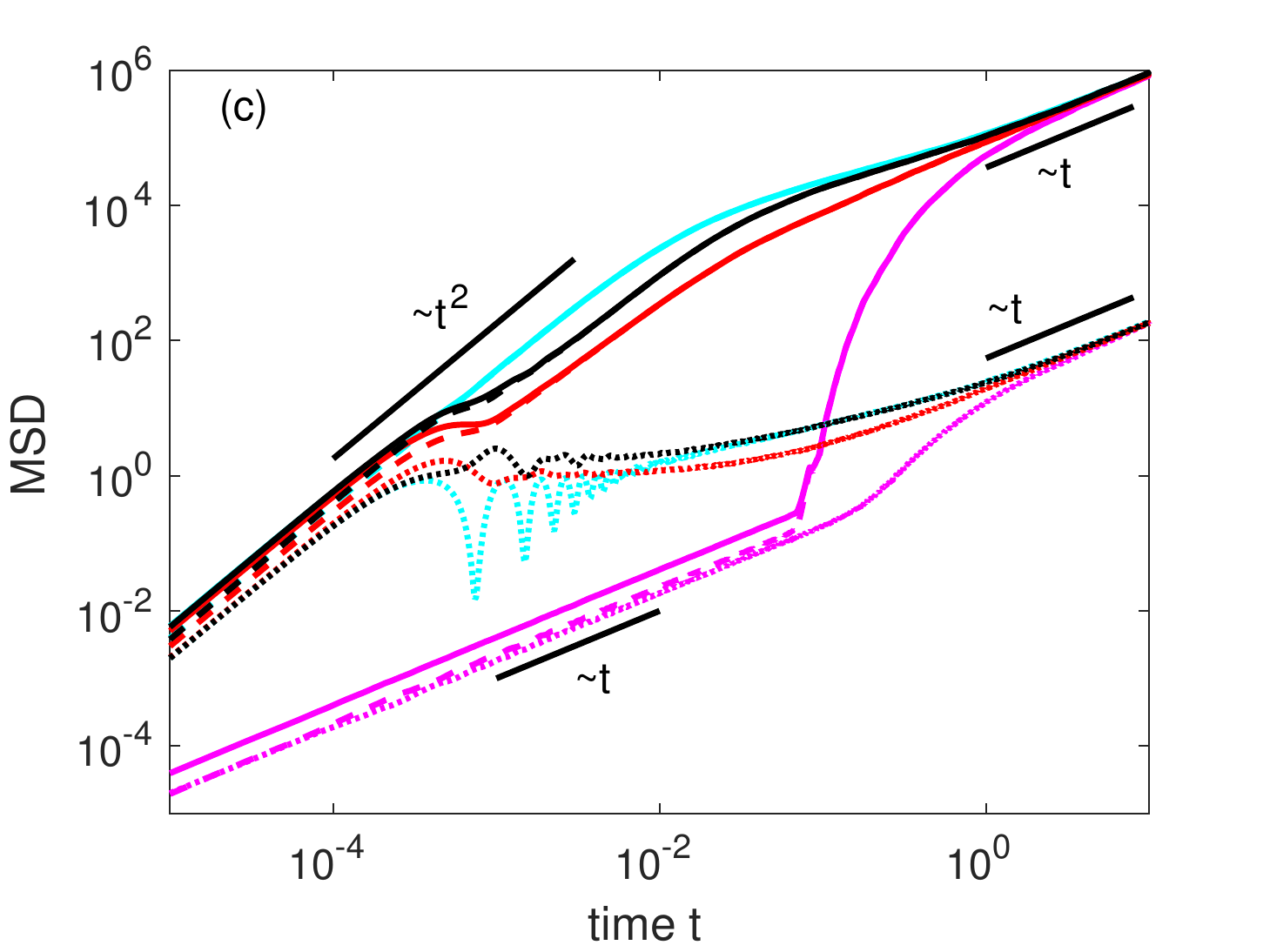}
\includegraphics[width=80mm]{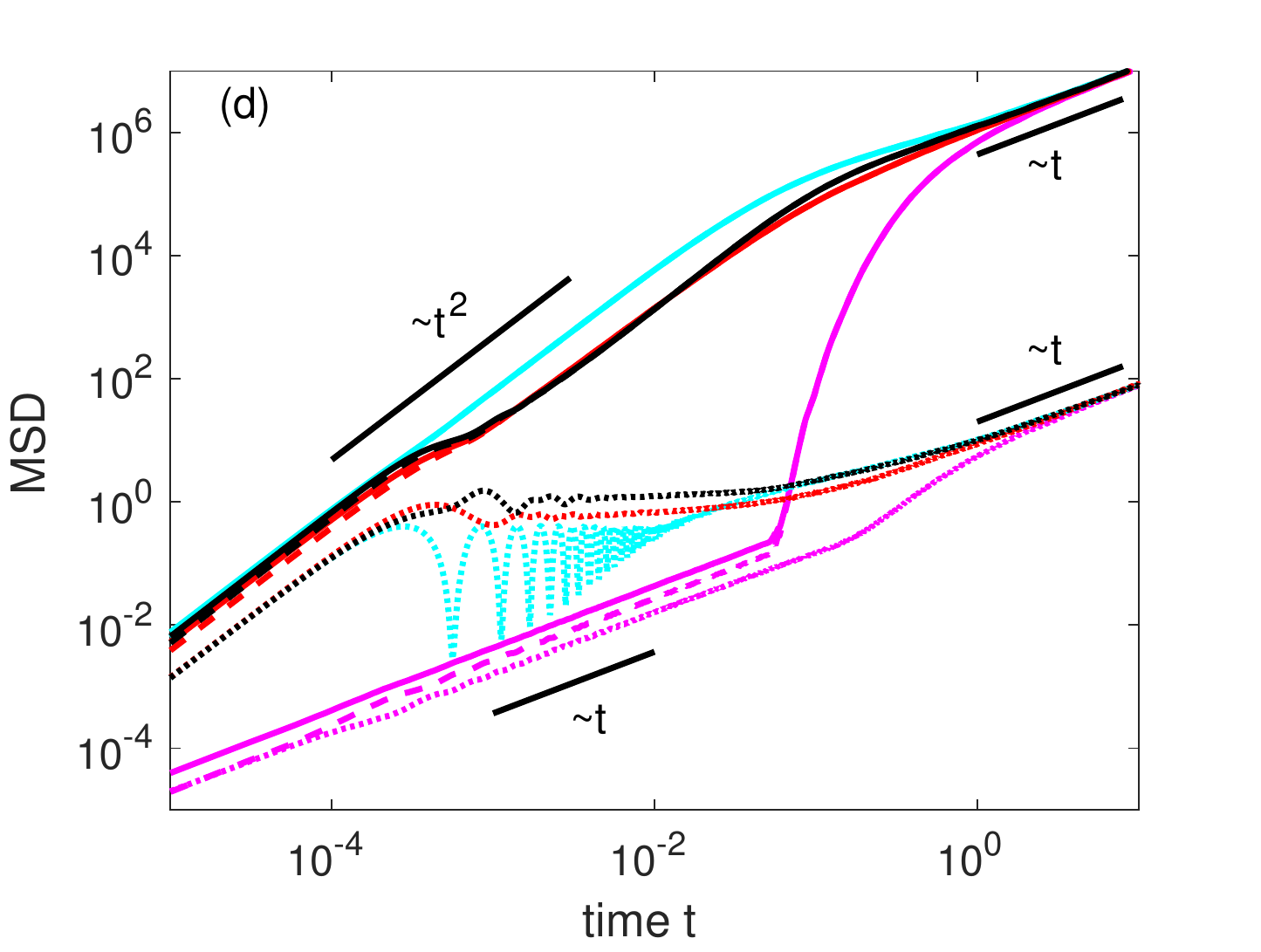}
\caption{\label{MSD1}Temporal evolution of MSD for $10^4$ walks 
at $\mathrm{Pe}=10^4$. Solid lines: total MSD, dashed lines: MSD parallel to the jet,
dotted lines: MSD orthogonal to the jet. Starting positions are denoted
by coloring: central streamline of a jet (light cyan), center of an eddy 
(magenta), flooded (dark red) and the separatrix between jet and eddy (black).
Panels: 
(a) $A=10^{-3}$, (b) $A=10^{-2}$, (c) $A=10^{-1}$, and (d) $A=0.25$.  
Note that in (a) and (b) the jet region is so thin that the MSD for the first and
last initial condition (cyan and black) almost coincide. For (a) the
parallel and the perpendicular components of the MSDs are almost equal, and
the eddy lattice flow \cite{PoeschkeEtAl2016} is being reproduced. Note also
that in (d) the MSD for the flooded case 
(red) and the start on the separatrix (black) are very similar.}
\end{figure*}

\begin{figure*}[t] \centering
\includegraphics[width=80mm]{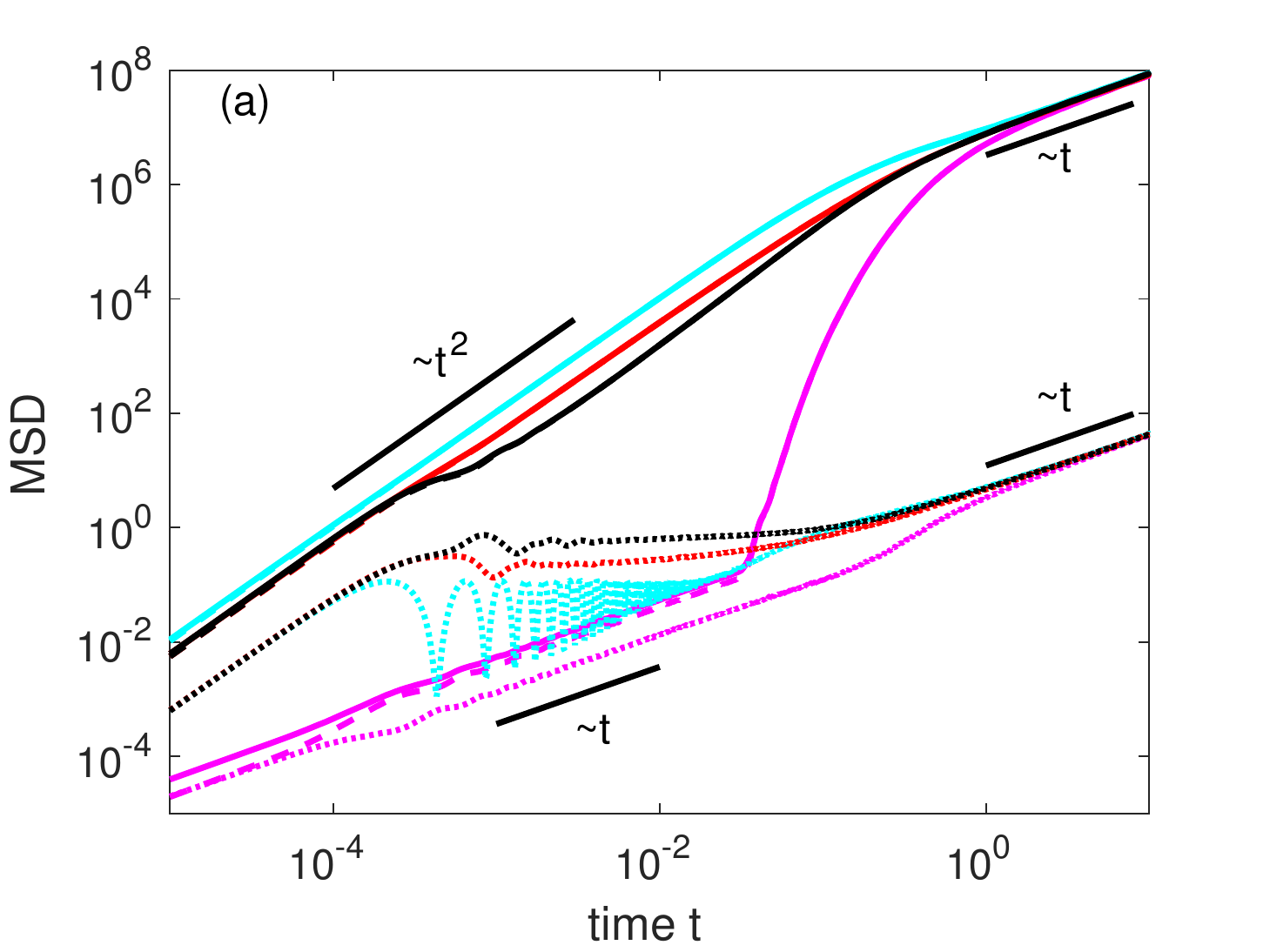}
\includegraphics[width=80mm]{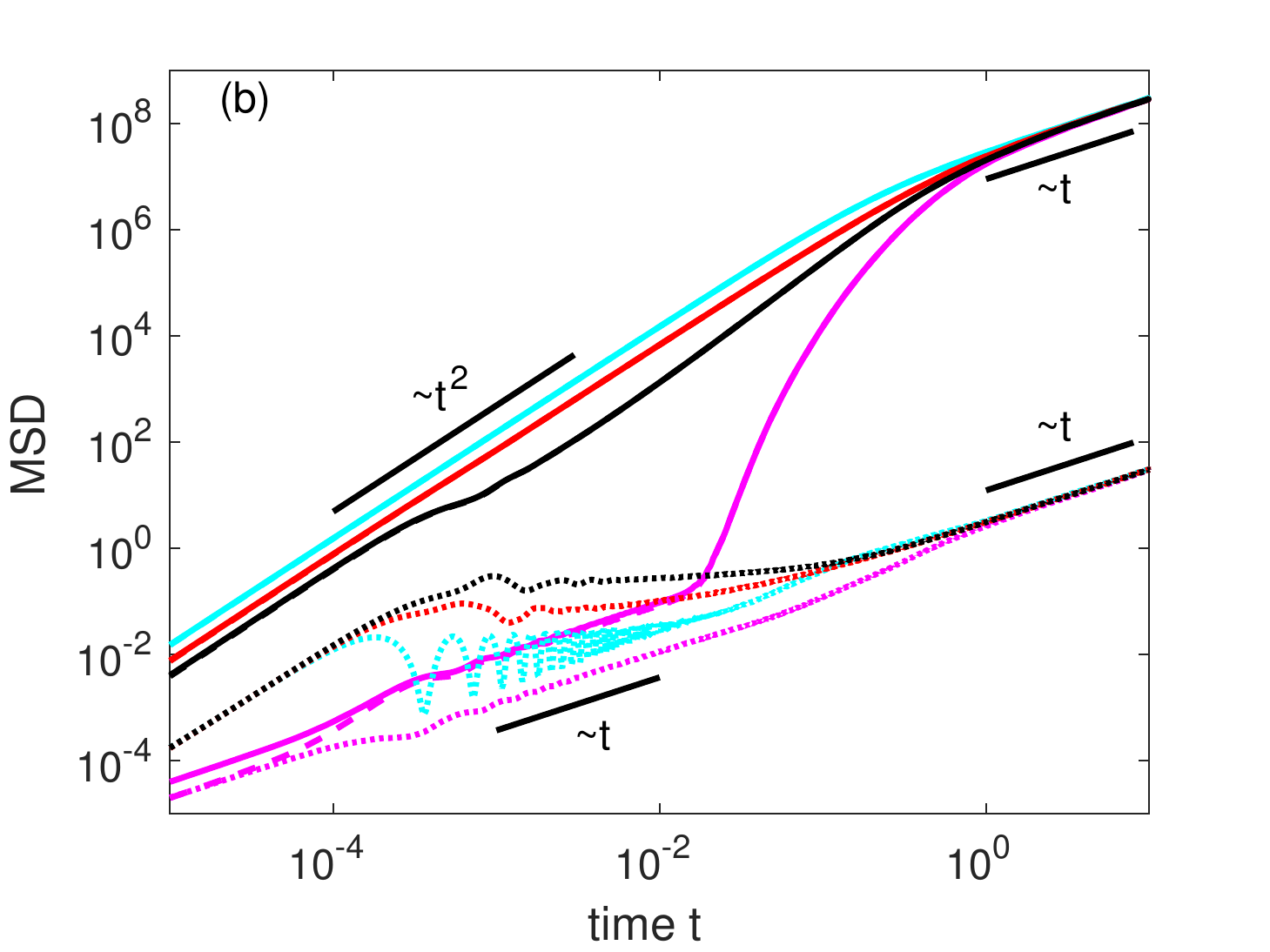}
\includegraphics[width=80mm]{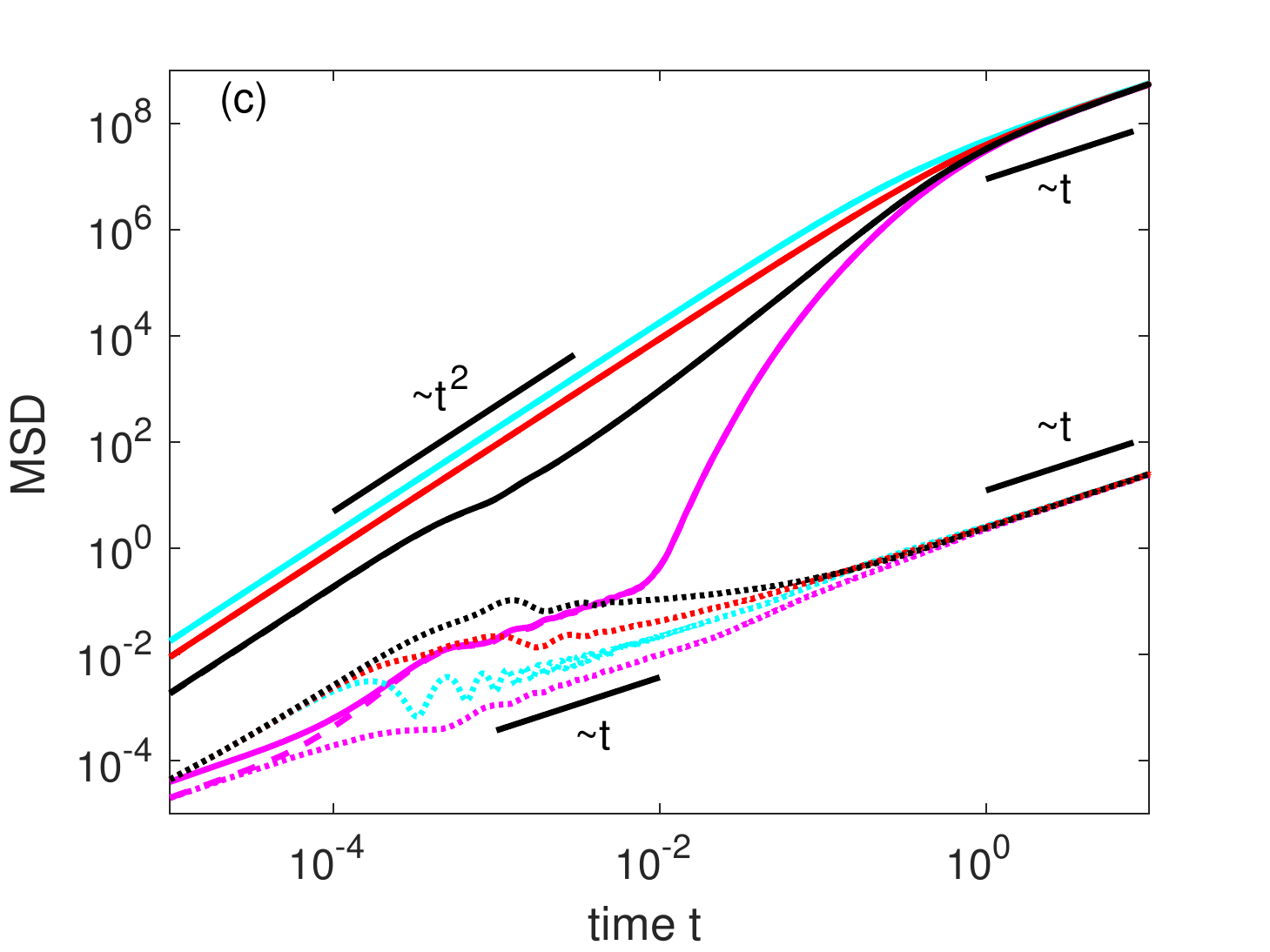}
\includegraphics[width=80mm]{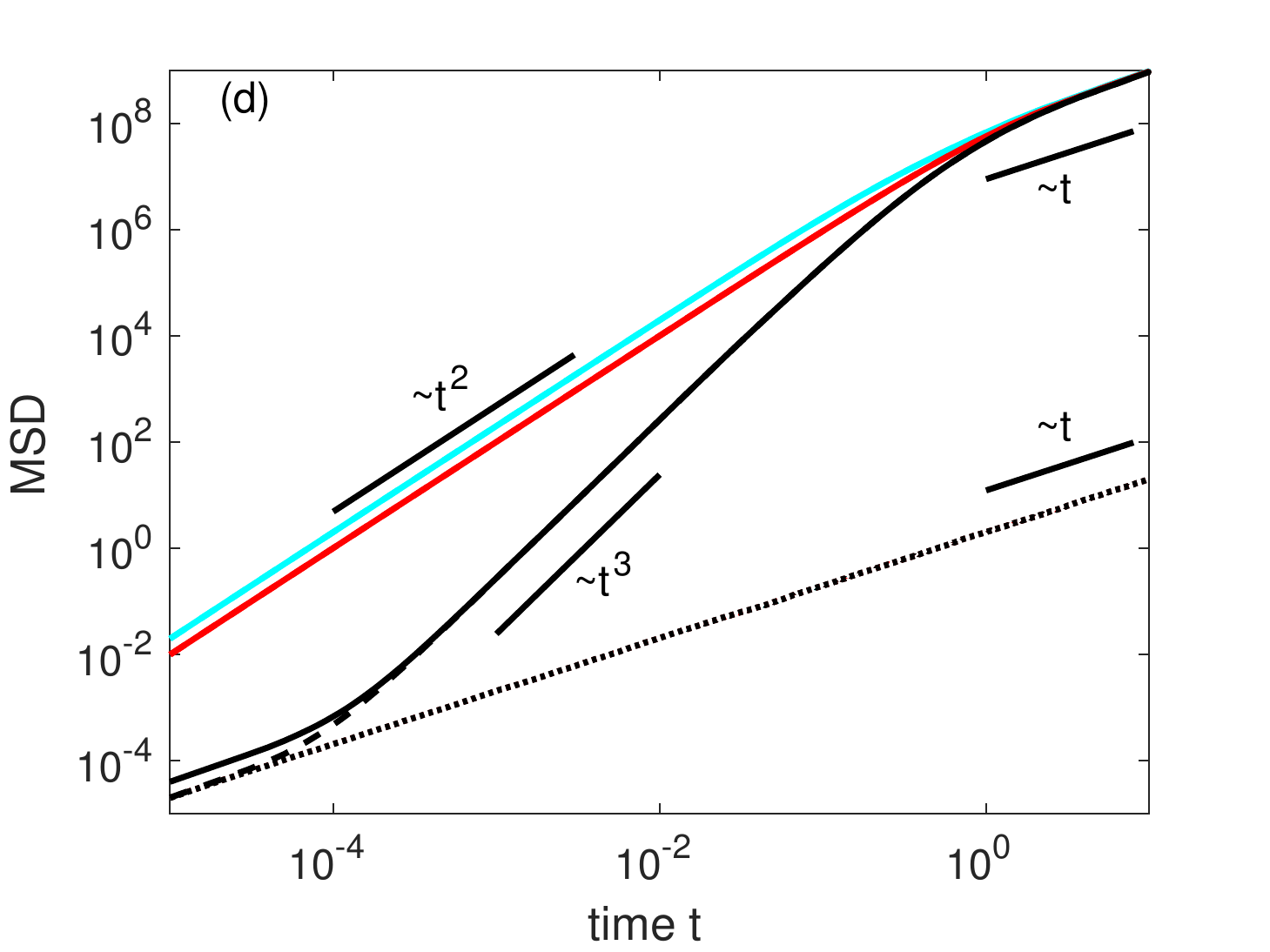}
\caption{\label{MSDlast}Same as Fig.~\ref{MSD1} with (a) $A=0.5$, 
(b) $A=0.75$, (c) $A=0.9$, and (d) $A=1$. 
Note that in the flow pattern of (d) there are no eddies, and 
the corresponding initial condition has converged 
to the one for a start at the separatrix (black).}
\end{figure*}

For the other examined initial conditions the MSD is very similar, except for
a start at the center of a cat's eye, see Figs.~\ref{MSD1} and \ref{MSDlast}.
For this initial condition an initial regime of normal diffusion turns 
after a short super-ballistic transient into a final regime of normal diffusion. Additionally, there is an intermediate regime of normal diffusion,
for moderate values of $A$, i.e. if $ 0.25\leq A \leq 0.9$. For most initial
conditions at not too small $A$ the overall MSD is almost identical to MSD
parallel to the axis: the MSD perpendicular to the axis is negligible. Both
components of the MSD display normal diffusion for times $t\gg t_2$. 
Our simulations confirm the observation that the corresponding 
final diffusion coefficients $D_\parallel^*$ and $D_\perp^*$ indeed 
possess functional dependences derived above, see \cite{Fannjiang} 
and Section~\ref{sect_MSD}. The simulations
indicate that these relations hold not only for $A\ll 1$ but approximately 
also in the broader range $A\leq 0.5$.

When the parameter $A$ approaches unity, the flow turns 
into a shear flow with a sinusoidal velocity profile. 
At $A=1$, the equations of motion in terms of the coordinates 
$x_+=x+y$ and $x_-=x-y$ become
\begin{eqnarray} 
\dot{x}_+ &=& \mathrm{Pe}\sin x_-\,+\,\sqrt{2}\,(\xi_x+\xi_y)\\
\dot{x}_- &=& \sqrt{2}\,(\xi_x-\xi_y). 
\end{eqnarray}
In the stripe where $\sin x_-\approx x_-$
this is approximately the linear shear flow \cite{Novikov}. 
The MSDs along both coordinates are well known
\cite{Foister} and read in our notation (note that $D=1$) 
\begin{eqnarray}
\label{MSDshear} \mathrm{MSD}_\parallel = \frac{1}{2}\left\langle x_+^2\right\rangle &=&
\frac{8}{3}\mathrm{Pe}^2 t^3\\ 
\mathrm{MSD}_\perp = \frac{1}{2}\left\langle x_-^2\right\rangle &=& 2t. 
\end{eqnarray} 
Indeed, numerics show that for $\mathrm{Pe}=10^3$ to $10^5$ the
MSDs are well fitted by  
$\mathrm{MSD}_\parallel^{\mathrm{sim}}=2\mathrm{Pe}^2t^3$ and
$\mathrm{MSD}_\perp^{\mathrm{sim}}=2t$. 
This means that for $A=1$ the system is close to a linear shear flow.

Noteworthy, for tracers starting close to the separatrix at $A\to 1$, the final normally diffusive regime 
is preceded by a superballistic one: a transition towards the shear flow $\mathrm{MSD}_\parallel \propto t^3$, see Eq.~(\ref{MSDshear}) and Fig.~\ref{MSDlast} (c,d), is being established.

\subsection{Aging}

\begin{figure*}[t] \centering
\includegraphics[width=80mm]{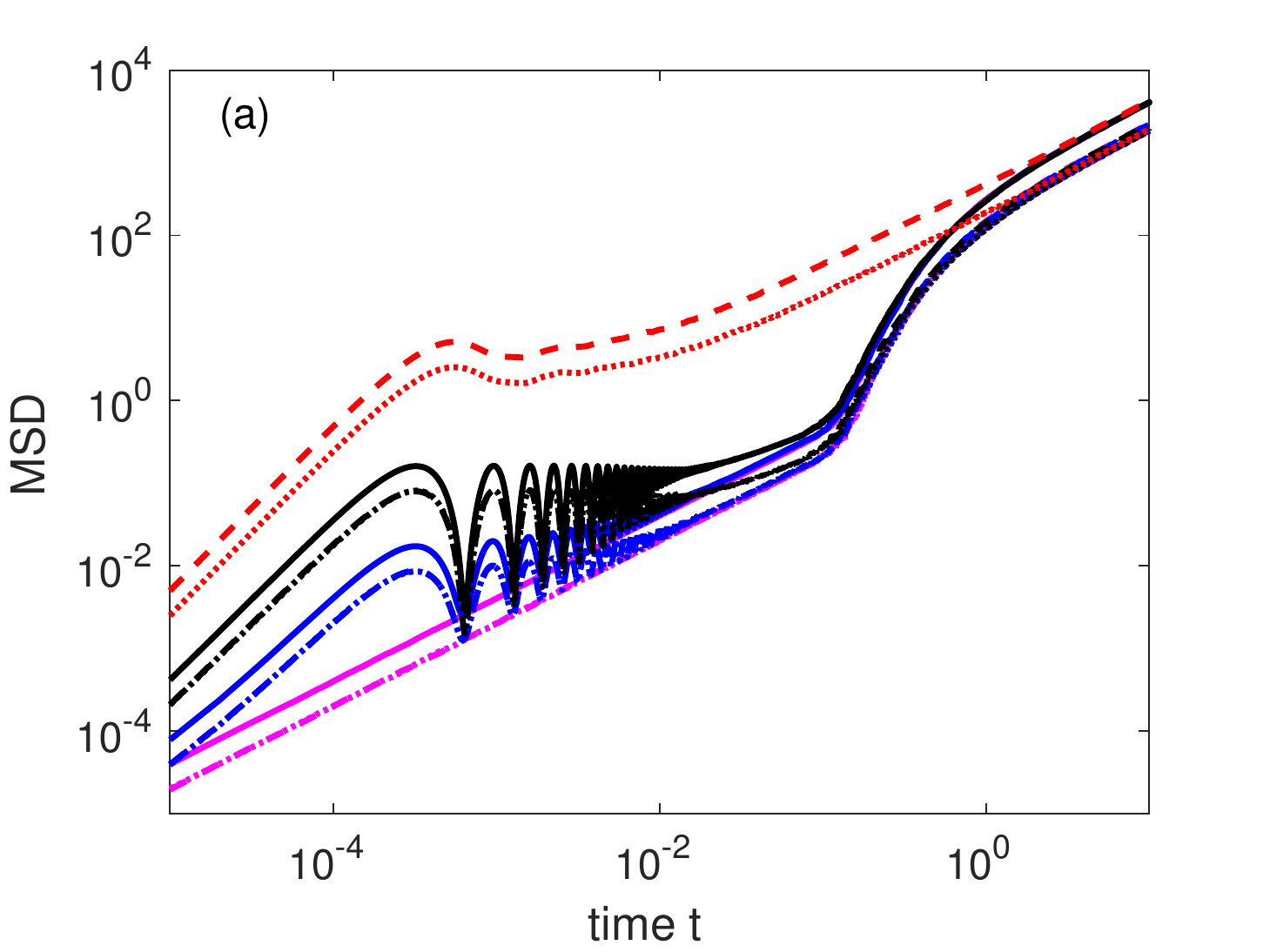}
\includegraphics[width=80mm]{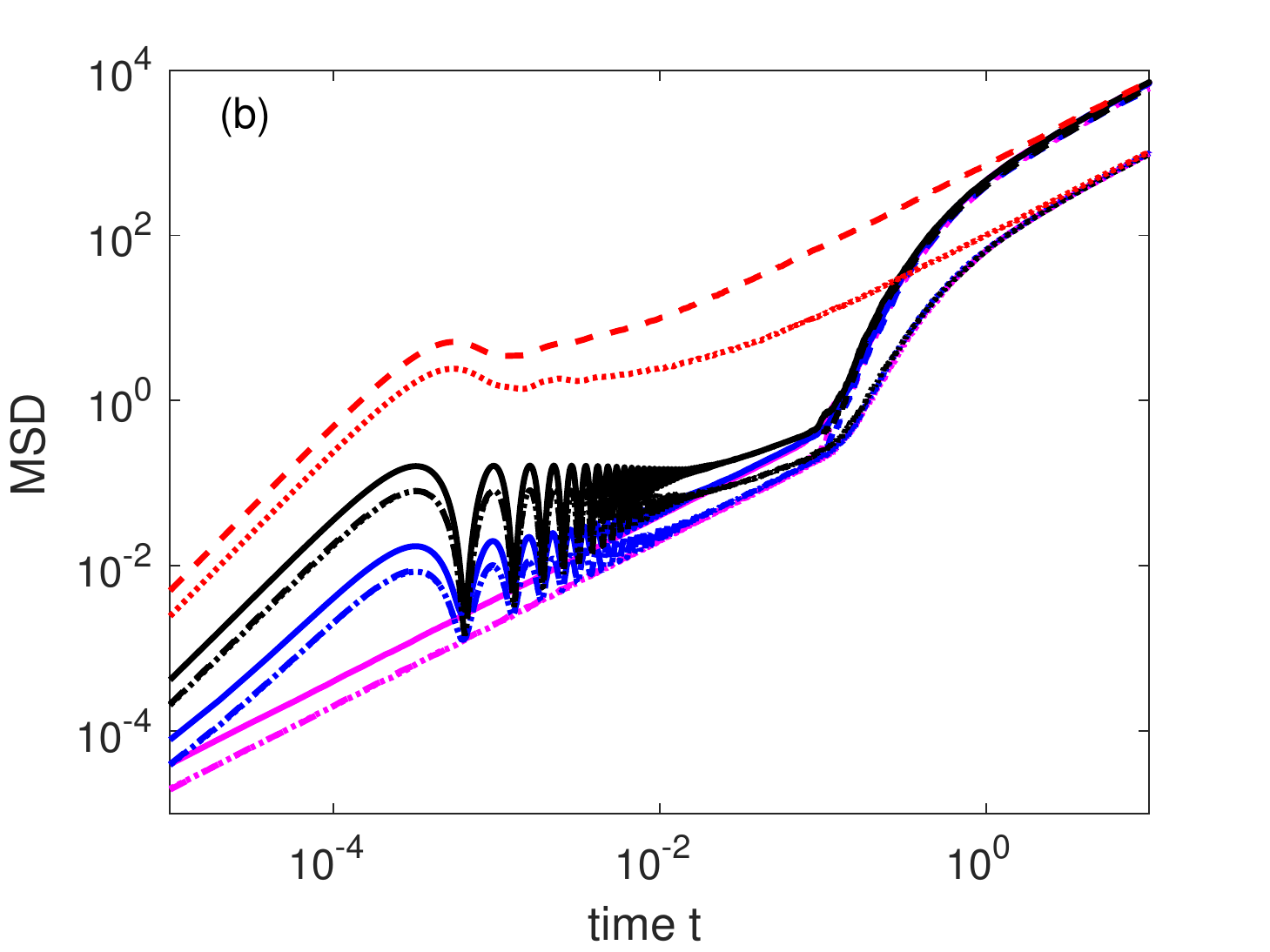}
\includegraphics[width=80mm]{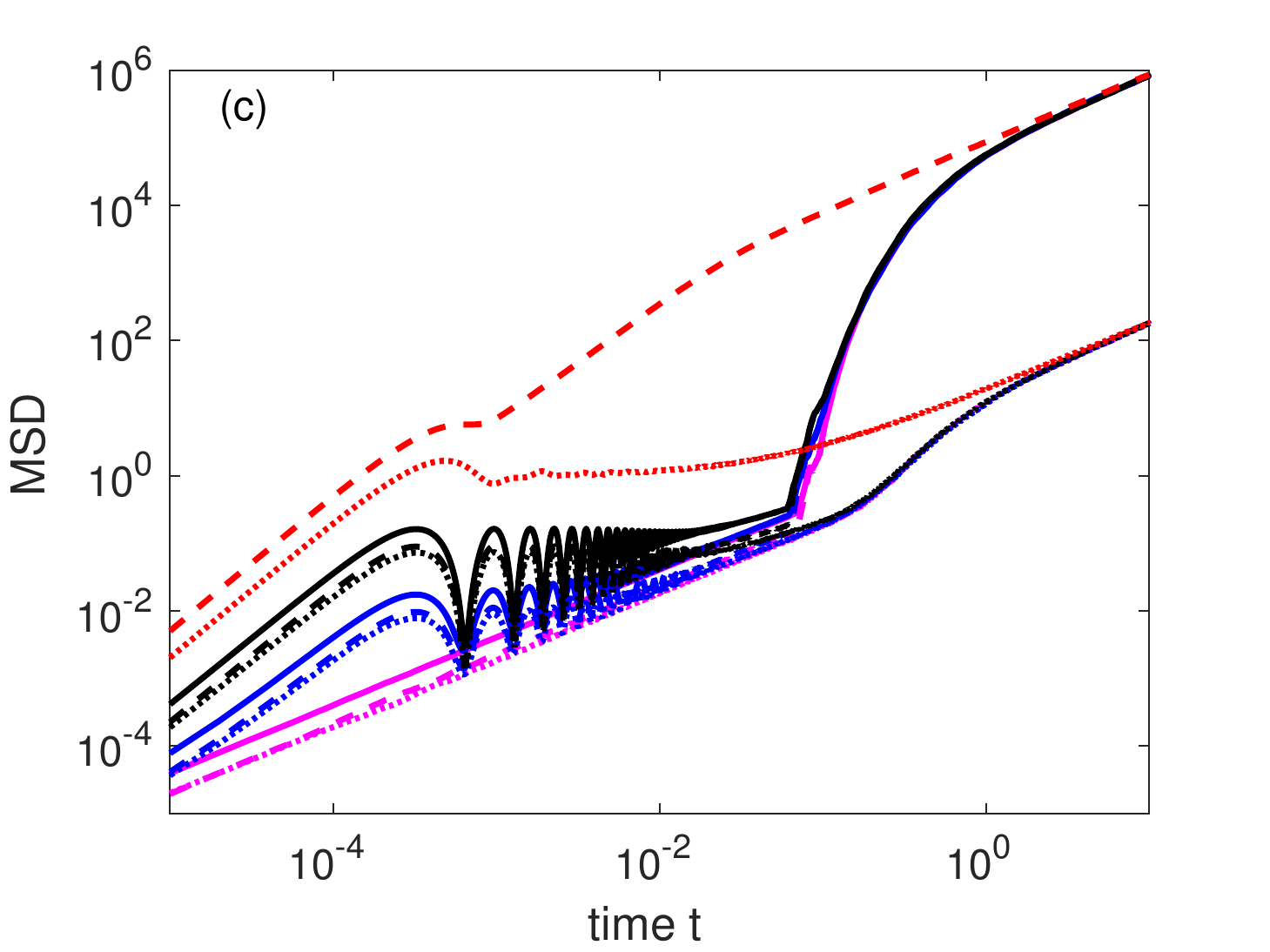}
\includegraphics[width=80mm]{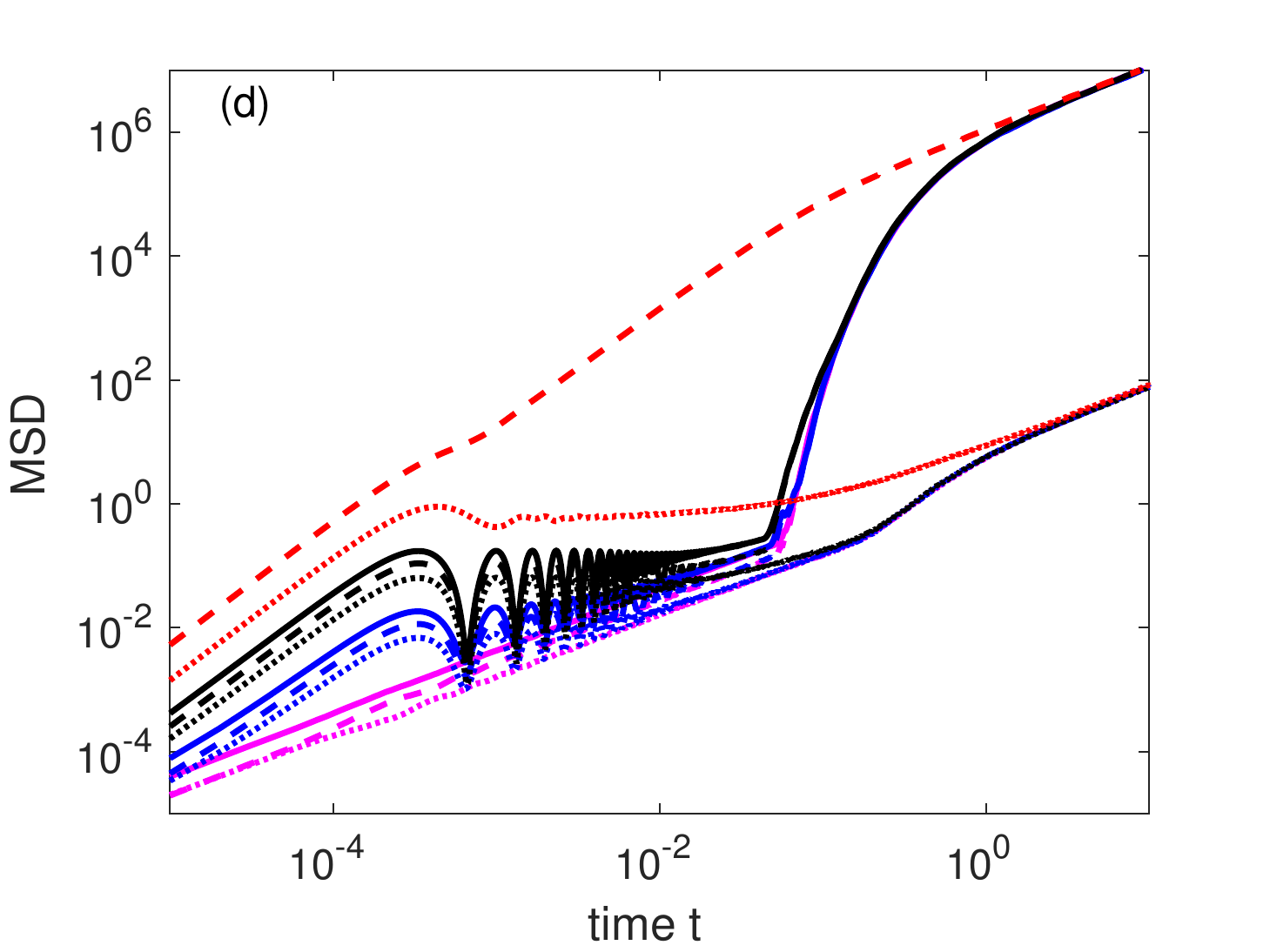}
\caption{\label{agedMSD1}
Aged MSD for $10^4$ walks starting at the eddy center at $\mathrm{Pe}=10^4$.
Solid lines: total MSD, dashed lines: MSD parallel to the jet,
dotted lines: MSD orthogonal to the jet. 
Aging times: $t_a=0$ (magenta), $t_a=10^{-3}$ (blue) 
and $t_a=10^{-2}$ (black).
Panels: (a) $A=10^{-3}$, (b) $A=10^{-2}$, (c) $A=10^{-1}$, 
and (d) $A=0.25$. 
For $t_a \gg t_2$ the MSD converges to the flooded case (red dashed) 
respectively to its orthogonal part (red dotted). 
In (a) the jet region is so thin, that
the eddy lattice flow is reproduced. Here the parallel and the orthogonal
components are approximately the same.} 
\end{figure*}

\begin{figure*}[t] \centering
\includegraphics[width=80mm]{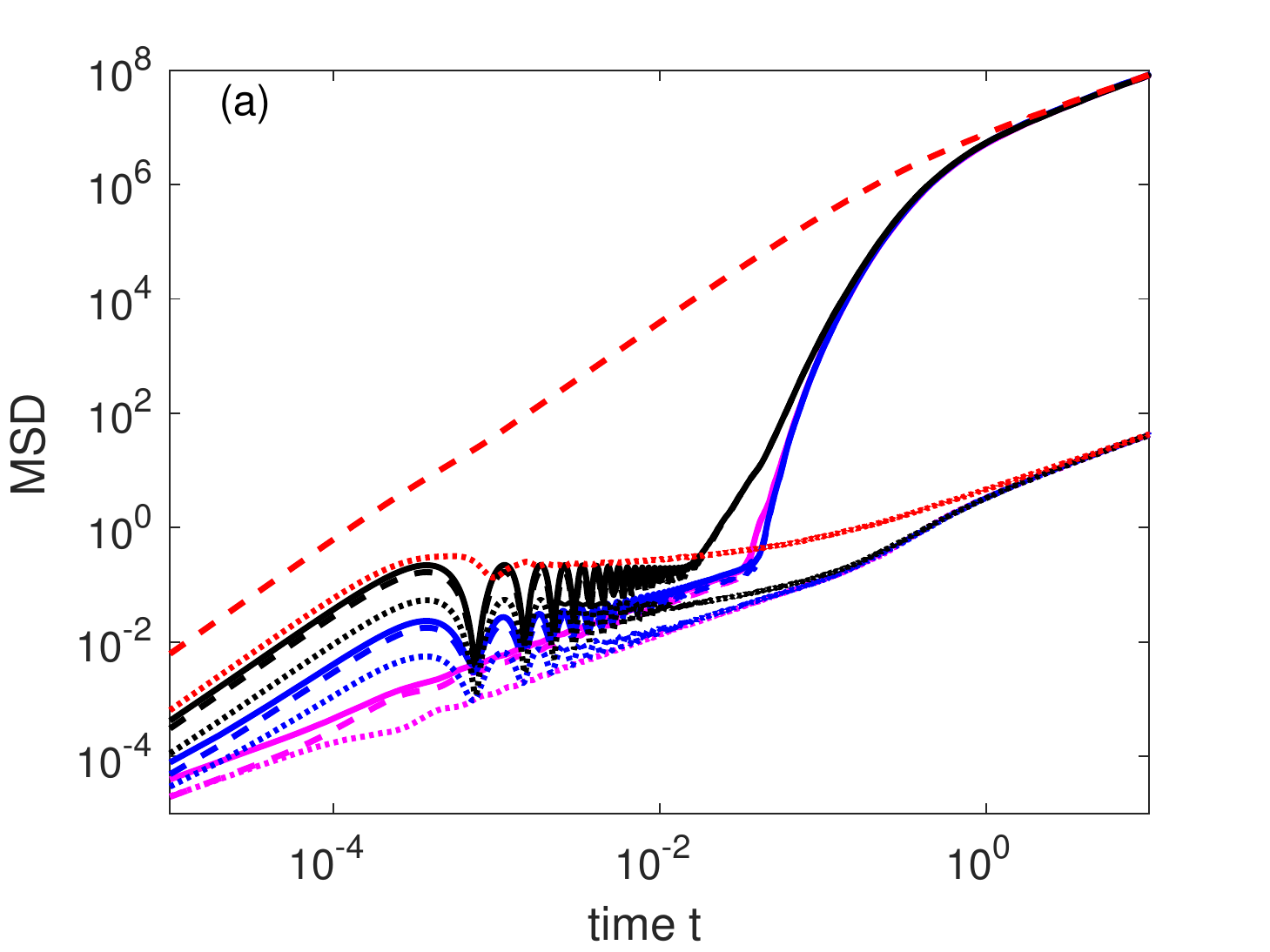}
\includegraphics[width=80mm]{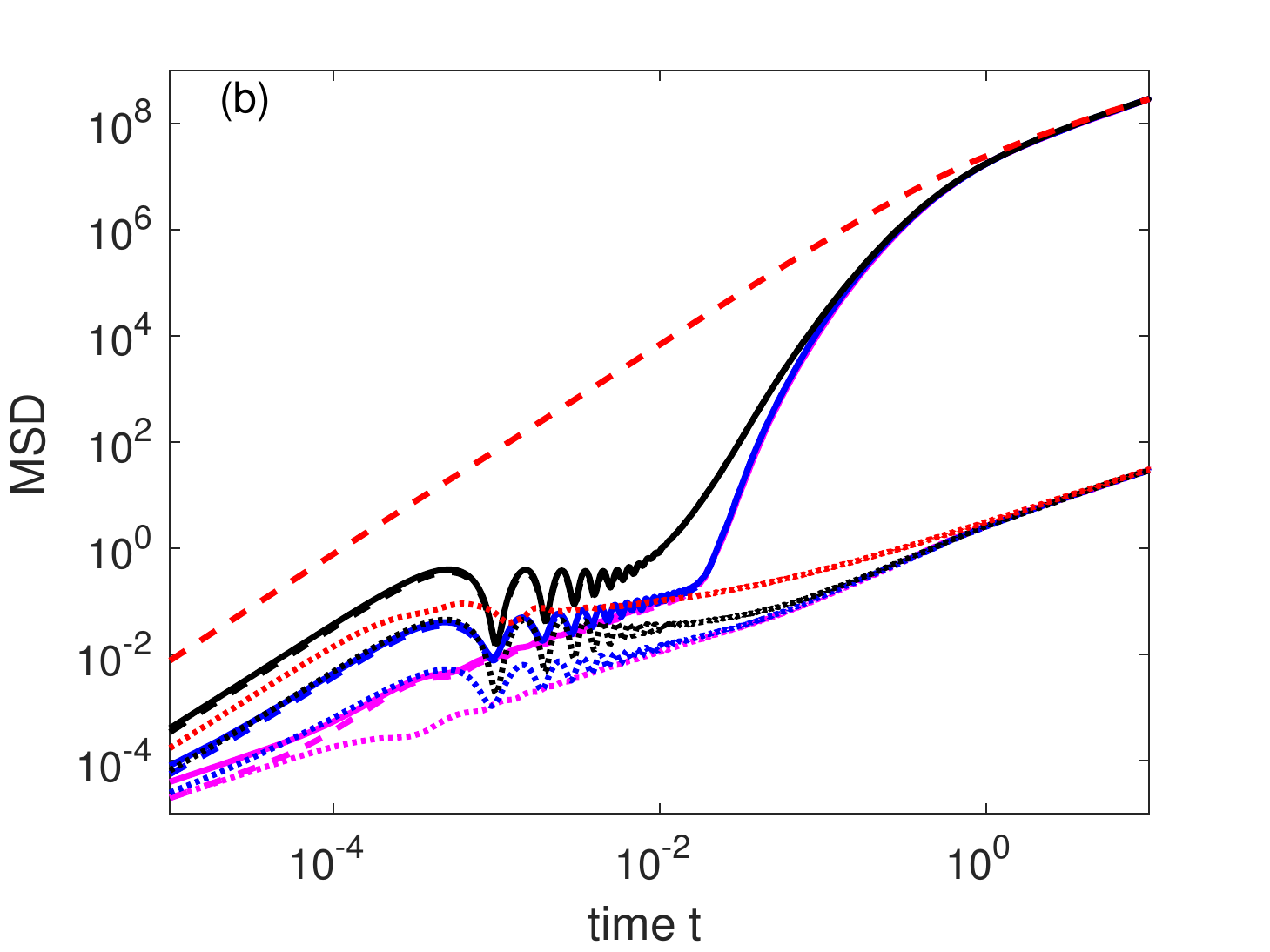}
\includegraphics[width=80mm]{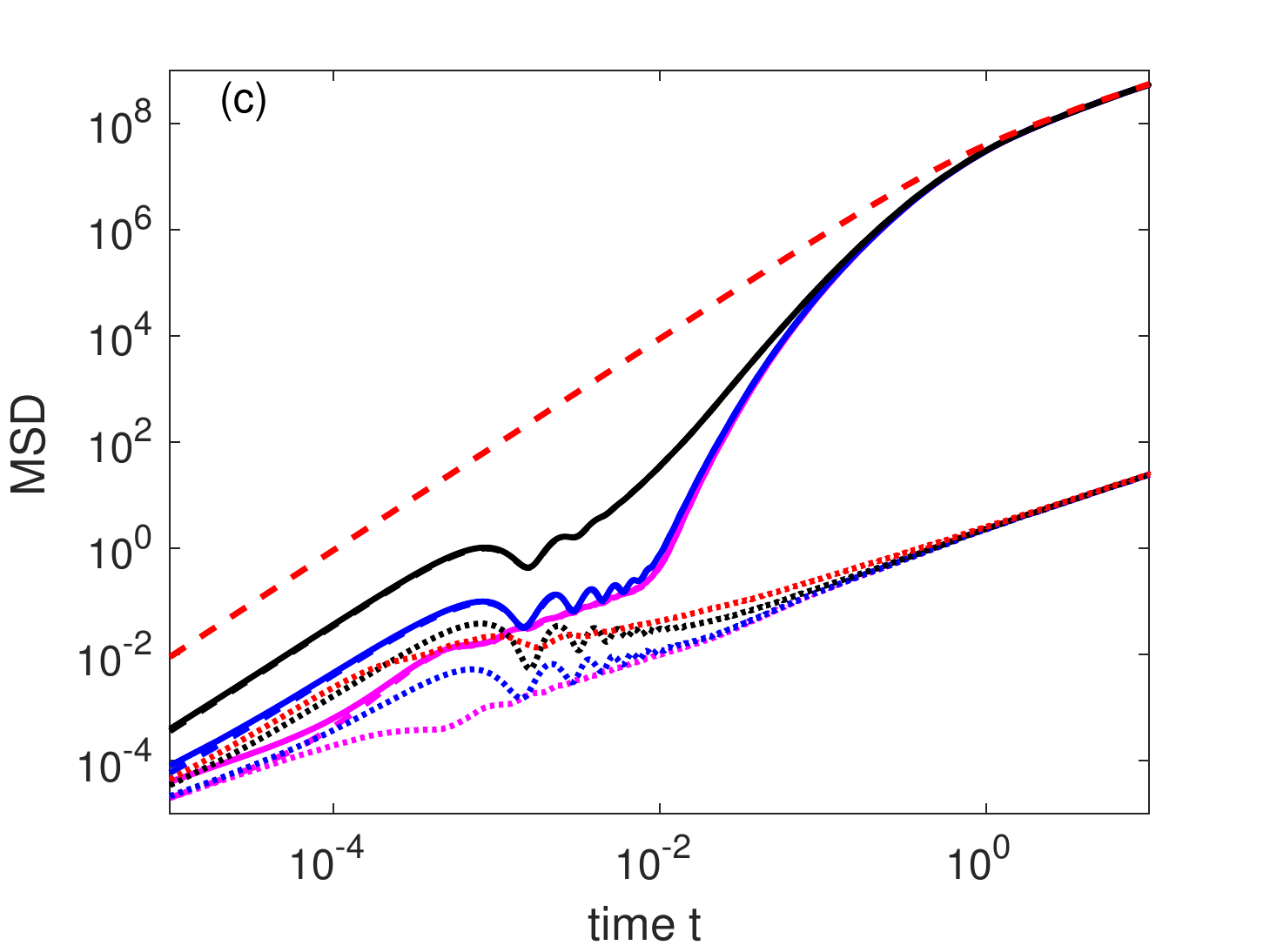}
\includegraphics[width=80mm]{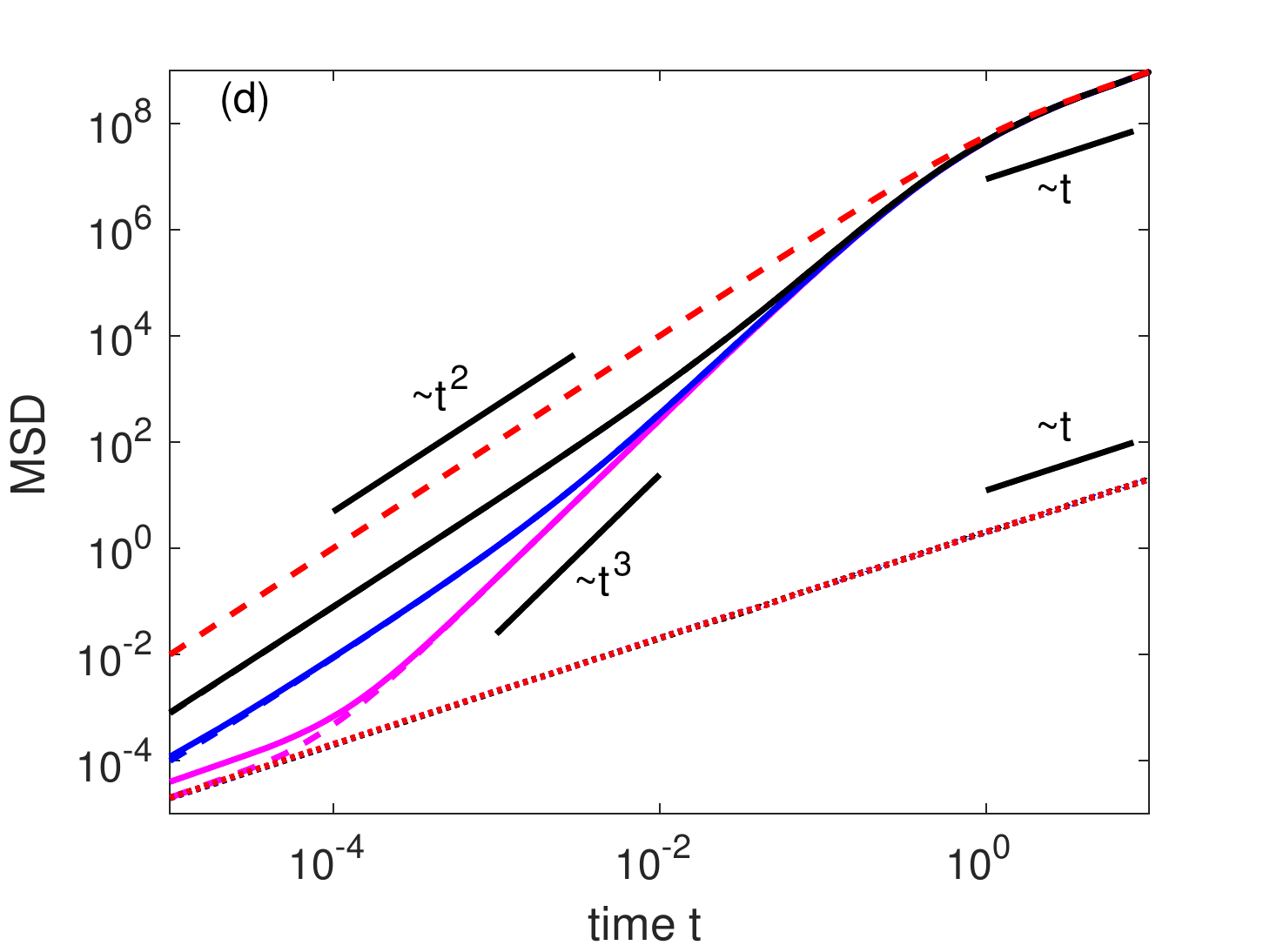}
\caption{\label{agedMSDlast}Same as Fig.~\ref{agedMSD1} with (a) $A=0.5$, 
(b) $A=0.75$, (c) $A=0.9$, and (d) $A=1$. 
Note that at $A=1$ there are no eddies. Their
former centers lie on the straight lines which consist of degenerate
equilibria. Here, aging for tracers that start on these lines is shown.}
\end{figure*}

For the initial position of tracers at the center of the cat's eye we also
considered the aged MSD: starting at $(x,y)=(\pi/2, \pi/2)$, letting the
tracers evolve for the aging time $t_a$, and then commencing the observation. 
Figures~\ref{agedMSD1} and \ref{agedMSDlast} show these aged MSDs. 
Note that the slopes are the same as in Figs.~\ref{MSD1} 
and \ref{MSDlast}. The MSDs for other initial conditions are already close 
to that for the flooded case and thus do age a lot less. 
Recall that for $A=1$ there are no eddies anymore; 
their former centers, as well as the former hyperbolic points, 
lie exactly on the separatrix which, in its turn, becomes a straight line
that entirely consists of degenerate stagnation points.
For this situation we show  in Fig.~\ref{agedMSDlast} (d) 
the process of aging  for starting on that straight line.

For $A\neq 1$ the aged MSD as a function of time is oscillating. 
As shown in \cite{PoeschkeEtAl2016}, for
sufficiently small $A$ and not too large values of times 
and aging times $t, t_a\ll 1$,
the leading terms for the aged MSD are given by
\begin{eqnarray} 
\mathrm{MSD}(t,t_a) &\approx & 8(t+t_a)+\frac{8(t+t_a)}{\left[1+(2\Omega\,t(t+t_a))^2\right]^2} \\
\nonumber && \times
\left\{\left[(2\Omega\,t(t+t_a))^2-1\right]\cos(\Omega\,t) \right. \\
\nonumber && \qquad \left. -4\Omega\,t(t+t_a)\sin(\Omega\,t)\right\},
\end{eqnarray} 
where the frequency of oscillations $\Omega\approx \mathrm{Pe} \,\sqrt{1-A^2}$
is a monotonically decreasing function of $A$.
For this approximation to work well, the aging time should strongly
exceed one period: $t_a\gg {2\pi}/{\Omega}$.

\section{Conclusions} 
\label{sect_Conclusion}

We considered the advection-diffusion problem for tracer particles in the
cat's eye flow.  This two-dimensional flow consists of jet regions having a shape of meandering strips,
and eddies having a shape of cat's eyes. The family of the cat's eye flows is parametrized by
a single parameter, and interpolates between the eddy lattice flow (without jets) and a shear flow 
with sinusoidal velocity profile (without eddies). In the absence of molecular diffusion, the tracers are either 
carried away ballistically by jets, or stay trapped in eddies. 
Adding small thermal noise makes possible the transitions between eddies and jets, and, at long times,
leads to anisotropic diffusion with the diffusion coefficient in the jet's direction much larger than 
the one in the perpendicular direction. This long time regime seems to be the only one which was 
discussed theoretically in considerable detail. 
At intermediate time scales the transport can be modeled by a stochastic scheme corresponding to 
L\'evy walks interrupted by rests. The transport phase of the walk corresponds to the motion in a jet, and 
rests to trapping in eddies. This scheme however only applies for the initial conditions
corresponding to starting close to the separatrix. The behavior for other initial conditions 
may be vastly different. 

In the present work we provide results of extensive numerical simulations of the particles' transport by the 
cat's eye flow concentrating on the mean squared displacement of the particles from their initial positions
in a broad time domain, and investigate its intermediate-time behavior, the influence of initial conditions, and aging regimes, i.e. the behavior of 
MSD between some intermediate time $t_a$ and final observation time $t > t_a$. The results of simulations confirm theoretical 
results for the long time behavior of MSD, and the applicability of the L\'evy walk scheme for intermediate times, including the 
prediction about the connection between the transport velocities in the short- and intermediate-time ballistic
regimes. They also show a multitude of possible aging behaviors (depending on initial conditions), including an oscillatory
one which is observed when particles start inside the eddies. This oscillatory behavior is 
due to the particles' rotation in an eddy during the trapping phase, and is not captured by the L\'evy walk scheme. 

\section*{Acknowledgments}

This work was financed by the German-Israeli Foundation for Scientific Research
and Development (GIF) grant no. I-1271-303.7/2014.

\appendix

\section{Derivation of the intermediate asymptotic MSD} \label{sect_B}

In this section we derive the asymptotic expression for the MSD,
Eq.~(\ref{Second_ballistic}), for intermediate times when starting at the separatrix. 
Since the MSD is dominated by the longitudinal motion along the axis of the system, a one-dimensional model
is adequate. We use {\it L\'evy walks interrupted by rests} as a theoretical description. 
The derivation follows~\cite{KlaS}.

Let $P_1(x,t)$ be the probability density of a tracer being at position $x$ at
time $t$ when starting in a ballistic mode and alternating between ballistic
motions with velocity $v$ and rests. 
Given the probability densities of waiting times inside a jet
(without index), for resting times (with index $r$) as well as for the last,
incomplete step respectively rest (upper-case symbols), we obtain

\begin{eqnarray} 
P_1 (x,t) &=& \Phi(x,t)+\int_0^t \phi(x,t')\Phi_r (t-t')\, dt'\\ 
\nonumber && +\int_{-\infty}^\infty dx' \int_0^\infty dt' \int_0^{t'} dt'' \\
\nonumber && \times \phi(x',t'')\phi_r (t-t')\Phi(x-x',t-t')+\dots,
\end{eqnarray} 
respectively 
\begin{eqnarray} 
P_1(k,s) &=& \Phi(k,s) + \phi(k,s)\Phi_r (s)  \\ 
\nonumber && + [\phi(k,s)\phi_r (s)]^1 \Phi(k,s) \\
\nonumber && + [\phi(k,s)\phi_r(s)]^1 \phi(k,s)\Phi_r(s) + \dots \\ 
\nonumber && + [\phi(k,s)\phi_r (s)]^n \Phi(k,s) \\ 
\nonumber && + [\phi(k,s)\phi_r(s)]^n \phi(k,s)\Phi_r(s) + \dots 
\end{eqnarray} 
in Fourier-Laplace representation. By applying the geometric series 
to odd and even terms separately and averaging the
result with the one from an analog calculation for starting in the resting
phase, we arrive at 
\begin{equation} 
\label{PDFtheory} 
P(k,s) =
\frac{\Phi(k,s)(1+\phi_r(s))+\Phi_r(s)(1+\phi(k,s))}{2(1-\phi_r (s)\phi(k,s))}
\end{equation} 
for the probability density of being at time $t$ at position $x$
on the axis Fourier-transformed in space and Laplace-transformed in time.
Numerics show that the waiting time densities of a tracer inside a jet
respectively a vortex can roughly be approximated by a power law 
$\propto t^{-3/2}$, if the parameter $A$ is not too small, 
as expected in theory. 
Hence we get 
\begin{eqnarray} 
\phi_r(s) &=& 1-\sqrt{\tau s} \\
\Phi_r(s) &=& \sqrt{\frac{\tau}{s}} \\
\Phi(k,s) &=& \mathrm{Re}\left\{\sqrt{\frac{\tau}{s+ivk}}\right\} \\
\nonumber &=& \frac{\sqrt{\tau}\cos\left(\frac{1}{2}
\arctan\left(\frac{vk}{s}\right)\right)}{(s^2+v^2 k^2)^{1/4}}
\end{eqnarray} 
as well as
\begin{equation} 
\phi(k,s) = \frac{1}{2}\sqrt{\frac{\tau}{\pi}} \int_{\tau/\pi}^\infty t^{-3/2}\cos(kvt)e^{-st}\; dt. 
\end{equation}
Writing the cosine complex one obtains 
\begin{eqnarray} 
\phi(k,s) &=& e^{-\frac{\tau}{\pi}s}\cos\left(\frac{\tau}{\pi}kv\right) \\ 
\nonumber && +
\frac{\sqrt{s+ikv}}{2}\left[-\sqrt{\tau}
+\mathrm{erf}\left(\sqrt{\frac{\tau}{\pi}(s+ikv)}\right)\right]\\ 
\nonumber && + \frac{\sqrt{s-ikv}}{2}
\left[-\sqrt{\tau}+\mathrm{erf}\left(\sqrt{\frac{\tau}{\pi}(s-ikv)}\right)\right].
\end{eqnarray} 
Substituting everything into (\ref{PDFtheory}) and expanding both
its numerator and its denominator separately first in $k$ until second order 
and then in $s$ until first order, 
keeping only the highest order terms in $s$ in each coefficient 
of the series of $k$ yields 
\begin{equation} 
P(k,s) = \frac{4\sqrt{\frac{\tau}{s}}-\frac{3\sqrt{\tau}v^2}
{4s^{5/2}}k^2}{4\sqrt{\tau s}+\frac{\sqrt{\tau}v^2}{4s^{3/2}}k^2} =
\frac{1}{s}\frac{\left[1-\frac{3}{16}\left(\frac{vk}{s}\right)^2\right]}
{\left[1+\frac{1}{16}\left(\frac{vk}{s}\right)^2\right]},
\end{equation} 
i.e. 
\begin{equation} 
P(k,s) = \frac{1}{s}\left[1-\frac{1}{4}\left(\frac{vk}{s}\right)^2\right] 
\end{equation}
in second order in $k$. From the probability density follows the MSD
\begin{equation} 
\mathrm{MSD}(s) = \frac{1}{2}v^2 s^{-3}, 
\end{equation} 
which according to Tauberian theorems for the inverse
Laplace-transform corresponds to (\ref{Second_ballistic}) in real time.

\end{document}